\pdfoutput=1
\documentclass[usenatbib]{mn2e}
\usepackage{apjfonts}
\usepackage{amssymb}
\usepackage{amsmath}
\usepackage{ctable}
\usepackage{fixltx2e} %% forces figure order to remain fixed (otherwise figure*'s can move out-of-order)

\newcommand{\be}{\begin{equation}}
\newcommand{\ee}{\end{equation}}

\newcommand{\msun}{M_{\sun}}

\newcommand{\scaleup}{}

\newcommand\plotone[1]
 {\centering \leavevmode \includegraphics[width={0.99\columnwidth}]{#1}}

\newcommand{\plotside}[1]
 {\centering \leavevmode \includegraphics[width={0.95\textwidth}]{#1}}

\newcommand{\acknowledgments}{\begin{small}\section*{Acknowledgments}\end{small}}
\newcommand\altaffilmark[1]{$^{#1}$}
\newcommand\altaffiltext[1]{$^{#1}$}
\title[Inflows in Eccentric Disks]{Origins, Structure, and Inflows of 
$m=1$ Modes in Quasi-Keplerian Disks}

\author[Hopkins]{
\parbox[t]{\textwidth}{ 
Philip F. Hopkins\altaffilmark{1}\thanks{E-mail:phopkins@astro.berkeley.edu}} 
\vspace*{6pt} \\
\altaffiltext{1}{Department of Astronomy, University of California
  Berkeley, Berkeley, CA 94720} \\
}

\date{Submitted to MNRAS, September, 2010}
\begin{document}
\maketitle
\label{firstpage}

\begin{abstract}
Simulations show eccentric disks ($m=1$ modes)
forming around quasi-Keplerian potentials, a topic of interest for 
fueling quasars, forming super-massive BHs, 
planet formation and migration, 
explaining the origin and properties of 
nuclear eccentric stellar disks like that in M31, and driving the formation of the 
obscuring AGN torus. 
We consider the global, linear normal $m=1$ modes in collisionless 
disks, without the restriction that the disk mass be negligible relative to the 
central (Keplerian) mass. We derive their structure and 
key resonance features, 
and show how they arise, propagate inwards, 
and drive both inflow/outflow and eccentricities in the disk. 
We compare with hydrodynamic simulations of such disks around a super-massive 
BH, with star formation, gas cooling, and feedback. 
We derive the dependence of the normal mode structure 
on disk structure, mass profiles, and thickness, and 
mode pattern speeds and growth rates. 
We show that, if the disk at some radii has mass of $\gtrsim10\%$ the central point mass, 
the modes are linearly unstable and are self-generating. They arise as 
``fast modes'' with pattern speed of order the local angular velocity at these 
radii. 
The characteristic global normal modes have pattern speeds
comparable to the linear growth rate, of order $(G\,M_{0}\,R_{0}^{-3})^{1/2}$, where 
$M_{0}$ is the central mass and $R_{0}$ is the radius where the enclosed 
disk mass $\sim M_{0}$. 
They propagate inwards by exciting eccentricities 
towards smaller and smaller radii, until at small radii these are ``slow modes.''
With moderate amplitude, the global normal modes can lead to shocks 
and significant gas inflows at near-Eddington rates 
at all radii inside several $\sim R_{0}$. 
\end{abstract}


\section{Introduction}
\label{sec:intro}

Perturbations to normally circular orbits in nearly-Keplerian potentials 
are of fundamental interest for a variety of topics in astrophysics. For example, 
questions related to the fueling of super-massive 
black holes (BHs), their accretion disks, and the dynamics of nearby 
systems, the formation and fueling of protostars, and the behavior of 
protoplanetary disks and planets around stars or rings and moons 
around planets. 
Particularly interesting are 
perturbations with azimuthal wavenumber $m=1$ 
(amplitude $\propto \cos{\phi}$), which can manifest as 
eccentric orbits or disks, lopsided or sloshing modes, or 
one-armed spirals. 
It is easy to see why: 
the response of a nearly circular orbit to a 
weak perturbation, to leading order, scales with 
$1/[\kappa^{2}-m\,(\Omega-\Omega_{p})^{2}]$, 
where $\Omega$ is the orbital frequency, 
$\kappa$ is the epicyclic frequency, and $\Omega_{p}$ is the 
characteristic frequency (precession rate) of the perturbation. 
In a Keplerian potential, $\kappa=\Omega \propto r^{-3/2}$, 
so (since $\Omega_{p}$ is finite) for any continuous system 
this scales at small radii as $\sim1/(1-m)\,\Omega^{2}$. For 
general $m$, this vanishes, but for $m=1$ the leading terms cancel 
and there is a strong resonant response. 
Physically, this 1:1 resonance between radial and azimuthal 
frequencies is related to the fact that elliptical orbits in a Keplerian 
potential are closed and do not precess. As a consequence, 
the eccentricity distribution and mode behavior in such a disk 
can be determined by collective effects in the disk, 
even where these collective effects are very weak compared to the 
gravity of the central object. 

Recently, for example, \citet{hopkins:zoom.sims} have shown that 
the formation of lopsided, eccentric disks within the BH radius of influence is a 
ubiquitous feature in hydrodynamic simulations of massive gas inflows 
in galaxies, and that such disks can efficiently drive gas angular momentum 
loss and power BH accretion rates of up to $\sim10\,\msun\,{\rm yr^{-1}}$. 
The co-existence of gas and stars is critical for the large inflow rates seen; 
the torques on the gas are dominated by the mode in the collisionless portion of the disk. 
And the stellar relics of these disks bear a remarkable similarity to 
nuclear disks observed on $\lesssim10\,$pc scales around 
nearby supermassive BHs, particularly the well-studied case at the center of 
M31 \citep{lauer93}, whose origin has been mysterious. 
The inflow and outflow regulated by these modes in such simulations also determines 
the nature of the galaxy mass profile on scales within $\lesssim10-100\,$pc. 
As such, it is also particularly interesting to understand whether or not such 
modes could arise generically.
%, and if so could 
%explain the characteristic power-law like slopes observed 
%at the centers of ``cusp'' ellipticals. 
There are many candidate nuclear disks in such systems \citep{lauer:ngc4486b,
  lauer:centers,
  houghton:ngc1399.nuclear.disk,
  thatte:m83.double.nucleus,debattista:vcc128.binary.nucleus,
  afanasiev:2002.ngc5055.nuclear.disk,
  seth:ngc404.nuclear.disk}. 
%And cuspy galaxy centers are in fact observed to be disky, with high 
%ellipticities, and probably reflect the nature of inflow at the last major gas-rich 
%event in the nucleus \citep{faber:ell.centers,ferrarese:type12,jk:profiles}. 
%But their characteristic profile shapes have largely been unexplained -- 
%at the smallest radii ($<1\,$pc) they can be explained by relaxation forming a Bahcall-Wolf 
%cusp \citep[the relaxation time being $\gg t_{\rm Hubble}$ at larger 
%radii;][]{bahcallwolf}, and at larger radii ($\gtrsim100\,$pc) by inflows generated by 
%mergers or violent instabilities \citep{hopkins:cusps.mergers,hopkins:cusps.ell}, 
%but no good explanation yet exists for the behavior in between.

There is considerable 
literature discussing the mode structure, pattern speeds, and evolution of 
general self-gravitating disk instabilities \citep[see e.g.][]{lin.shu:spiral.wave.dispersion,
goldreichtremaine:spiral.excitement,goldreichtremaine:spiral.resonances,
toomre:spiral.group.velocity,toomre:spiral.structure.review}. 
But these $m=1$ modes are less well-understood, especially in 
collisionless (stellar or planetary) disks. 
For example, there remains considerable debate regarding the 
stability of such modes \citep[e.g.][]{tremaine:slow.keplerian.modes,
salow:nuclear.disk.models,jacobs:longlived.lopsided.disk.modes,
touma:keplerian.instabilities}. 
These works describe many interesting behaviors of 
$m=1$ modes in a disk in a nearly Keplerian potential, 
but their conclusions rely on specific assumptions. 
\citet{tremaine:slow.keplerian.modes} show that such modes 
are linearly stable, but only in the limit where 
$M_{d}\ll M_{\rm BH}$ at all radii. 
And much of the insight from the study of protostellar and planetary disks 
focuses on either the same limit (for planetary disks) or 
the opposite limit (for early-stage protostellar disks) in which the system 
is really just a self-gravitating disk 
\citep[see][and references therein]{laughlin:1994.protostar.disk.instabilities,
laughlin:1996.protostar.disk.instabilities.inflow}
But in simulations or observations of 
galactic nuclei \citep{levine2008:nuclear.zoom,
  escala:nuclear.gas.transport.to.msigma,mayer:bh.binary.sph.zoom.sim,
  hopkins:maximum.surface.densities,hopkins:zoom.sims}
and protostellar evolution \citep{laughlin:1994.protostar.disk.instabilities,
bate:1995.protobinary.accretion.vs.frag,
nelson:1998.circumstellar.disk.instabilities}, 
much of the interesting behavior involves disks over a range of 
radii where the disk is not completely negligible in mass.
The mode structure outlined in \citet{tremaine:slow.keplerian.modes} 
is used in \citet{papaloizou:2002.m1.modes} to estimate 
the effects on a one or two low-mass planet system, but assuming 
mode stability and under similar mass and radius 
restrictions (and not allowing for a large collisionless disk component). 
\citet{adams89:eccentric.instab.in.keplerian.disks}
and \citet{ostriker:eccentric.waves.via.forcing,shu:gas.disk.bar.tscale} 
reach opposite conclusions 
to \citet{tremaine:slow.keplerian.modes}, but they focus on 
``fast modes'' ($\Omega_{p}\sim \Omega$) in the regime $M_{d}\sim M_{\rm BH}$ 
(or $M_{d}\sim M_{\ast}$), 
and consider only pure fluid disks with a hard and reflecting outer edge, in which case 
the resulting mode growth rates depend sensitively on the structure and nature of 
reflection at the disk edge (and can be dramatically modified -- in fact, 
completely eliminated -- allowing any 
flow ``through'' or outside the edge, the most probable configuration). 
In general, collisionless disks can have very different mode structure 
from fluid systems. 
\citet{christodoulou:fast.mode.torii.stability} consider instabilities of narrow torus annuli 
(as opposed to disks) but with similar restrictions. 
Also, because the mode growth is sensitive to motion of the center of mass, 
most numerical studies to date have lacked the resolution 
to determine whether mode growth is real or artificial in the limit where the disk 
mass is smaller than the BH/star mass 
\citep[see the discussion in][]{nelson:1998.circumstellar.disk.instabilities}.

Moreover, these studies have largely been restricted to very specific 
mass profiles (e.g.\ the Kuz'min disk, with $\Sigma\rightarrow$constant 
at small radii) and/or systems with sharp ``edges,'' 
where in fact the profiles seen in interesting 
phases in simulations and in e.g.\ star-forming systems and 
the centers of real ``cusp'' elliptical galaxies 
resemble a range of power-law slopes with smooth declines 
at large radii and significant variation in profile shape 
\citep{gebhardt96,
hopkins:msigma.scatter,hopkins:sb.ir.lfs}. 

As such, the origin of these modes, where observed, and 
many of their properties have remained ambiguous. 
Given claims of the stability of such modes, alternative suggestions for their 
origin have ranged from 
their being induced by an external collision/passage 
\citep[in galactic nuclei, from e.g.\ a nuclear star cluster; see][]{sambhus:m31.nuclear.disk.model}, 
to their being excited by substantial populations of stars on retrograde 
orbits \citep{touma:keplerian.instabilities}. 
But these explanations pertain only to very specific 
systems or regimes, and do not explain the ubiquity of 
such modes in astrophysical systems. Showing that they can in fact 
be self-generating via gravitational instability would have profound implications.

Moreover, in systems with star or planet formation as an ongoing 
process, these modes will evolve in time, as quantities such as the 
disk gas fraction, mass profile, dispersion profile, 
and total disk mass are affected by these processes. 
Similarly, the modes in the collisionless disk can drive very strong inflows and 
outflows in gas, changing the properties of the 
host disks in turn \citep[see][]{nelson:1998.circumstellar.disk.instabilities,
bournaud:2005.lopsided.disk.inflow,
hopkins:zoom.sims}. Therefore it is of great importance to both survey a 
range of analytic profile shapes and dispersion levels, 
and to compare with simulations that can incorporate non-linear effects 
on the mode evolution. It is also particularly interesting to examine the 
level of inflow generated by such modes, if the disks have some gas. 

In this paper we expand upon the investigation of global $m=1$ modes in 
nearly Keplerian, predominantly 
collisionless disks. In particular, we focus on the question of whether 
or not such modes can, in fact, be unstable and self-generating, and 
if so how they propagate throughout such disks. 
After defining some terms (\S~\ref{sec:definitions}), we 
consider modes in the local (WKB) limit (\S~\ref{sec:wkb}), which allows us to 
analytically derive approximate stability criteria and 
discuss conditions for efficient mode propagation, in both gaseous and 
stellar disks. 
In \S~\ref{sec:exact}, we discuss exact numerical solutions which allow us to 
extend this discussion to linear global normal modes, and outline under what conditions 
these modes are unstable (and what their characteristic frequencies are), 
and how this depends on a variety of disk 
properties including mass profile shape, mass, and 
disk thickness. We discuss the structure of such 
modes, and the conditions under which they will drive shocks in collisionless+gaseous 
systems and corresponding inflow/outflow, and the radii over which the 
modes can act. 
In \S~\ref{sec:sims}, we compare to the results from high-resolution hydrodynamic 
simulations which include self-gravity, gas cooling, star formation, shocks, 
inflow/outflow, and non-linear effects all not captured in an analytic 
formulation, and discuss how this compares to our analytic insight. 
Finally, in \S~\ref{sec:discussion}, we summarize our conclusions 
and their implications for astrophysical systems.

\section{Definitions}
\label{sec:definitions}

Adopt a cylindrical coordinate frame, 
($R$, $\phi$, $z$), and consider an initially axisymmetric, thin, 
planar disk with an 
arbitrary spherical (BH+bulge+halo) component. 
The coordinate center is located at the BH.

The initial potential in the disk plane can be 
written $\Phi_{0} = \Phi_{0}(R)$, and other properties 
are defined in standard terms: 
\begin{align}
\Phi_{0} &= \Phi_{0}(R) \\
V_{c}^{2} & = R\, \frac{\partial \Phi }{\partial R} 
\approx \frac{G\,M_{\rm enc}(<R)}{R} \\ 
\Omega & \equiv  t_{\rm dyn}^{-1} = V_{c}/R \\ 
\kappa^{2} &\equiv  R\,\frac{{\rm d} \Omega^{2}}{{\rm d} R} + 4\,\Omega^{2} = 
\frac{\partial^{2}\Phi}{\partial R^{2}}+3\,\Omega^{2}
\end{align}
where $V_{c}$ is the circular velocity, $\Omega$ the angular 
velocity, and $\kappa$ the epicyclic frequency. 
We use $c_{s}$ to denote the sound speed in a gaseous 
disk and $\sigma_{z}$ the vertical dispersion in a stellar disk.

Consider a perturbation in the plane and define the perturbed 
surface density field by 
\begin{equation}
\Sigma \rightarrow \Sigma_{0}(R) + \Sigma_{1}(R,\,\phi) 
\end{equation}
We consider a frame rotating with the perturbation pattern 
speed $\Omega_{p}$. 
We can represent the perturbed system as a sum of linearly 
independent modes $m$, 
$\Sigma_{1}\equiv \sum_{m=1}^{\infty}\,\Sigma_{m}$. 
Since we will focus on the behavior of these modes individually, 
consider (without loss of generality), the case of a single mode, 
\begin{align}
\Sigma_{m} &\equiv \Sigma_{a}(R)\,\exp{\left\{ i\,(m\,\phi-\omega\,t)\right\}} \\ 
\Sigma_{a}(R) &\equiv |a(R)|\,\Sigma_{0}(R)\,\exp{\left\{i\,\int^{R}\,k\,dR \right\}}
\end{align}
where $m$ is the azimuthal wavenumber, $|a|=|a(R)|$ the effective mode amplitude, 
$k$ the radial wavenumber, and the complex $\omega$ the mode 
frequency. With these definitions, the mode pattern speed is 
$\Omega_{p}\equiv {\rm Re}(\omega)/m$, and the linear mode 
growth rate is $\gamma\equiv {\rm Im}(\omega)$ ($|a|\propto \exp{(+\gamma\,t)}$). 

The mode of particular interest is an $m=1$ mode, in a 
quasi-Keplerian potential. 
Simulations in which these inflows form self-consistently, including 
gas inflow, star formation, and feedback, typically find that the nuclear 
stellar and gas distributions are well-approximated by power laws on the 
scales of interest \citep{hopkins:zoom.sims}. 
We will therefore adopt a true power-law disk as a convenient reference model. 
For such a disk
\be
\Sigma\propto R^{-\eta} = \Sigma_{0}\,\left(\frac{R}{R_{0}}\right)^{-\eta}
\ee
and it is straightforward to show that 
\begin{align}
V_{c}^{2} &= 2\pi\,\alpha\,G\,\Sigma\,R \\ 
M_{d}(<R) &=2\pi\,(2-\eta)^{-1}\,\Sigma\,R^{2} \\ 
%\alpha &=-\frac{
%\Gamma[1-\frac{\eta}{2}]\,\Gamma[\frac{\eta-1}{2}]
%}{
%2\,\Gamma[\frac{3-\eta}{2}]\,\Gamma[\frac{\eta}{2}]
%}\,(1-\eta)
%\approx \eta
\alpha &=\frac{
\Gamma[1-\frac{\eta}{2}]\,\Gamma[\frac{1+\eta}{2}]
}{
\Gamma[\frac{3-\eta}{2}]\,\Gamma[\frac{\eta}{2}]
}
\end{align}
for $0<\eta<2$ ($\alpha\approx\eta$ for $\eta\le1$, $\approx1/(2-\eta)$ for $1\le \eta<2$). 
Around a central BH this gives 
\begin{align}
\Omega^{2} &=\frac{G\,M_{\rm BH}}{r^{3}}+
\frac{2\pi\,\alpha\,G\,\Sigma_{0}}{R_{0}}\,\left( \frac{R}{R_{0}} \right)^{-(\eta+1)} \\ 
\kappa ^{2} &=\frac{G\,M_{\rm BH}}{r^{3}}+
(3-\eta)\,\frac{2\pi\,\alpha\,G\,\Sigma_{0}}{R_{0}}\,\left( \frac{R}{R_{0}} \right)^{-(\eta+1)} \ .
\end{align}

We will also discuss finite disks, with a power-law cutoff (motivated by analogy to the 
Kuz'min disk) of the form 
\be
\Sigma= \Sigma_{0}\,\left(\frac{R}{R_{0}}\right)^{-\eta}\,
\left[1 + \left(\frac{R}{a}\right)^{2} \right]^{-(3-\eta)/2}
\label{eqn:sigma.cutoff}
\ee
so that $\Sigma\propto R^{-\eta}$ at small radii and $\Sigma\propto R^{-3}$ 
at large radii. The total mass in the disk is then 
\be
M_{d} = \Sigma_{0}\,a^{2}\,
\left(\frac{a}{R_{0}}\right)^{-\eta}\,
\frac{\pi^{3/2}\,\Gamma[1-\eta/2]}{\Gamma[(3-\eta)/2]}\ .
\ee
In this case, $\Omega$ (and the potential) 
must be evaluated numerically, but for purposes of interpretation, they 
can be approximated well (exactly at small/large $R$ and with $\sim10\%$ 
accuracy at $R\sim a$) by 
\be
\Omega^{2} \approx \frac{G\,M_{\rm BH}}{r^{3}} + 
\frac{2\pi\,\alpha\,G\,\Sigma_{0}}{R_{0}}\,\left( \frac{R}{R_{0}} \right)^{-(\eta+1)}\,
\left[1 + \mu\left( \frac{R}{a}\right)^{2} \right]^{-1+\eta/2}
\ee
where $\mu^{-(2-\eta)/2}\equiv (\pi^{1/2}\,\Gamma[1-\eta/2])/(2\,\alpha\,\Gamma[(3-\eta)/2])$. 
At small radii this is just the normal power-law disk $\Omega$, 
at large radii the disk portion simply becomes Keplerian, 
$\Omega^{2}\rightarrow G\,(M_{\rm BH}+M_{d})/r^{3}$. 

Although we, for convenience, define our canonical model with respect to a super-massive 
BH, many of our conclusions could identically be applied to {\em any} sufficiently 
collisionless disk 
in a quasi-Keplerian potential (e.g.\ protestellar or protoplanetary disks). 
In this case one should simply replace 
``black hole'' with ``central massive object'' (e.g.\ star), and 
stellar disk with whatever collisionless disk 
surrounds the system. 
Because of the scale-free nature of the above equations, 
the rescaling of units is trivial.

\section{The WKB Approximation}
\label{sec:wkb}

First, consider modes in the limit of the WKB approximation of 
tight-winding (i.e.\ local modes), where $|kR|\gg m$. We caution that this limit does not, in fact, 
hold for most of the global modes seen in simulations, but it is nevertheless 
instructive. 

\subsection{Instability Criteria}
\label{sec:instability}

The case of slow modes where the non-Keplerian part of the potential is everywhere 
small is presented in detail in \citet{tremaine:slow.keplerian.modes}. We briefly note 
some conclusions. 
Take the potential to be that of a central BH plus a non-Keplerian 
disk: 
\be
\Phi = \Phi_{\rm BH} + \Phi_{e} = - \frac{G\,M_{\rm BH}}{r} + \Phi_{e}
\ee
where $\Phi_{e}/\Phi_{\rm BH} \sim M_{\rm d}/M_{\rm BH}\ll 1$. 
Further, consider ``slow'' modes, where the pattern speed $\Omega_{p}\ll \Omega$. 
We can expand the equations of motion in parameters of 
$\mathcal{O}(\epsilon)$ where $\epsilon\equiv M_{\rm d}/M_{\rm BH}\ll1$. 
This gives the WKB dispersion relation (to leading order in $|kR|^{-1}$) of 
quasi-Keplerian slow modes, 
\be 
\omega = \varomega + \pi\,G\,\Sigma_{d}\,|k|\,\Omega^{-1} 
- c_{s}^{2}\,k^{2}\,\Omega^{-1}
\label{eqn:slowmode.dispersion.gas}
\ee
for a gas disk, 
or 
\begin{align}
\omega &= \varomega + \pi\,G\,\Sigma_{d}\,|k|\,\Omega^{-1} \mathcal{F} \nonumber \\
&\approx \varomega + \pi\,G\,\Sigma_{d}\,|k|\,\Omega^{-1} \,\exp{\left(-\beta\,|kR| \right)}
\label{eqn:slowmode.dispersion.stellar}
\end{align}
for a stellar disk, 
where we define 
\begin{align}
\nonumber 
\varomega &\equiv \frac{\Omega^{2}-\kappa^{2}}{2\,\Omega} \\ 
&= - \frac{1}{2\,\Omega}\,\left( \frac{2}{r}\,\frac{d}{dr} + \frac{d^{2}}{dr^{2}} \right)\Phi_{d}\ .
\end{align}
In the dispersion relation for a stellar disk, $\mathcal{F}$ is the standard reduction 
factor \citep{binneytremaine}, and the latter equality is a convenient approximation 
for softened gravity, with $\beta \approx \sigma_{z}/V_{c}\approx h/R$ (the stellar 
disk scale height). 

It is obvious that all terms on the right-hand side of 
Equations~\ref{eqn:slowmode.dispersion.gas}-\ref{eqn:slowmode.dispersion.stellar}
are real; therefore, quasi-Keplerian, ``slow'' $m=1$ modes are 
stable at this order 
\citep[for a more rigorous derivation, see][]{tremaine:slow.keplerian.modes}. 

We have, however, made a major assumption, that the disk mass is 
{\em everywhere} much less than the BH mass (and the mode is 
everywhere slow). 
In generality, and to second-order in $|kR|^{-1}$, the WKB dispersion 
relation can be written (for a gas disk) 
\begin{align}
(\omega - m\,\Omega)^{2} = & \kappa^{2} + 
{\Bigl (}k^{2} +\frac{m^{2}}{r^{2}}{\Bigr )}\,c_{s}^{2}\,(1+\chi) \\ 
& - 2\pi\,G\,\Sigma_{d}\,{\Bigl (}k^{2} +\frac{m^{2}}{r^{2}}{\Bigr )}^{1/2}\,\,(1+\chi)
\label{eqn:dispersion}
\end{align}
where 
\be
\chi = \frac{2}{1+(k\,r/m)^{2}}\,
\left(\frac{1-s}{1+s} \right)\,;\ \ \ \ \ s \equiv \frac{\partial \ln V_{c}}{\partial \ln R}
\ee
\citep{lau:spiral.wave.dispersion.relations}
\footnote{
We follow \citep{lau:spiral.wave.dispersion.relations} 
keeping in-phase terms to second-order in $|kR|^{-1}$ because the 
modes of interest are global ones. 
One can think of this as accounting for the effective minimum 
wavenumber $m$ from the azimuthal wave, and including the 
enhancement $1+\chi$ where $\chi=\Gamma\,\sin{i}$ 
($\Gamma\equiv \partial \ln{\Omega}/\partial \ln{R}$ and 
$i$ is the arm pitch angle), namely the leading-order 
term of the swing amplifier at this order in the WKB approximation. 
} 

Whenever the right-hand side of Equation~\ref{eqn:dispersion} is negative, 
the modes are unstable and grow exponentially. 
Take the case of interest, a global $m=1$ mode in a relatively cold disk. 
For convenience take the limit $k=0$ and $c_{s}=0$; Equation~\ref{eqn:dispersion} 
becomes
%\be
%(\omega-\Omega)^{2}=\kappa^{2} - 2\pi\,G\,\Sigma\,(1+2(1-s)/(1+s))
%(\omega-\Omega)^{2}=\kappa^{2} - 2\pi\,G\,\Sigma\,(3-s)/(1+s)
%\ee
\be
\left( \frac{\omega}{\Omega} -1 \right)^{2} = 2\,(1+s) - 2\,\left(\frac{3-s}{1+s}\right)\,\tilde{f}_{d}
\ee
where
\be
\tilde{f}_{d}\equiv \frac{\pi\,G\,\Sigma}{\Omega^{2}\,R} \approx \frac{M_{d}(<R)}{M_{\rm enc}(<R)}
\ee
is roughly the disk mass fraction inside $R$. 
The RHS is negative for $\tilde{f}_{d} > (1+s)^{2}\,(3-s)^{-1}$. 
If the potential is near-Keplerian, then $s\sim-1/2$, so this 
just becomes $\tilde{f}_{d}\gtrsim 1/10$. 

In greater detail, 
consider the special case of a cold Mestel ($\eta=1$) disk 
around a BH, with an $m=1$ mode
and mass ratio enclosed in some radius $y\equiv M_{d}(<R)/M_{\rm BH}$; 
the full dispersion relation from Equation~\ref{eqn:dispersion} is then
\be 
\left( \frac{\omega}{\Omega} -1 \right)^{2} = 
\frac{1}{1+y}\,\left[  
1+2\,y - y\,(|kR|^{2}+1)^{1/2}\,(1+\chi) 
\right]
\label{eqn:mestel.modes}
\ee 
where for a local mode, $|kR|\gg1$ and $\chi\rightarrow0$, 
while for a global mode $|kR|\rightarrow0$ but 
$1+\chi\rightarrow (7+6\,y)/(1+2\,y)$. 
Global modes are formally unstable then for 
$y>(-3+\sqrt{17})/4\approx0.281$ ($\tilde{f}_{d}=0.11$), and 
local modes unstable for $|kR|>(1+2\,y)/y$.  
The solutions for arbitrary power-law disks $\eta$ are tedious, 
but for global modes can be well approximated by 
$y>0.07+0.09\,\eta+0.07\,\eta^{2}+0.04\,\eta^{3}$. 

More generally, for the power-law disk+BH and the stellar 
dispersion relation, the minimum radius at which instability appears 
(noting that the term $|kR|\,\exp{\{-\beta\,|kR| \}}$ is maximized 
for $|kR|=1/\beta$) is given by 
\be
\frac{M_{d}(<R)}{M_{\rm BH}} \ge \frac{\beta\,e}{(2-\eta)\,(1-\beta\,\alpha\,e\,(3-\eta))}\ .
\label{eqn:instab.criterion}
\ee
For small $\beta$ this is just $y\gtrsim 1.35\,\beta\,(1-\eta/2)^{-1}$. 
If the disk does not extend to these masses, then it will be everywhere 
locally stable. It is also immediately clear that if 
$\beta \ge 1/(\alpha\,e\,(3-\eta))$ ($\approx0.2-0.3$ for the 
interesting range of $\eta$), then the disk is everywhere locally stable independent 
of $M_{d}/M_{\rm BH}$ (this is just $Q\gtrsim1$). 

This instability criteria agrees well with what is seen in simulations; 
for the simulations discussed in \S~\ref{sec:intro}, the mode growth 
rate and maximum mode amplitudes are plotted as a function of 
$M_{d}/M_{\rm enc}$ at radii $\sim10\,$pc near the BH radius of influence 
in Figure~6 of \citet{hopkins:inflow.analytics} (see also Figure~12 of 
\citealt{hopkins:zoom.sims}). Around these values of $y$ or $\tilde{f}_{d}$, 
rapid growth rates for the $m=1$ mode appear at these radii. So at least 
at larger radii, mode growth is possible. 

What is the nature of these modes? 
Note that the right-hand side of Equation~\ref{eqn:mestel.modes} 
is real; as such, to lowest order in the WKB approximation, 
the unstable branch must correspond to an {\em overstability} 
with the real part of $\omega$, ${\rm Re}(\omega)=\Omega_{p}=\Omega$. 
In other words, the system can develop {\em fast} modes 
that are globally unstable, where $y$ is not very small. 

This suggests a picture in which the $m=1$ modes first appear at large 
radii -- some $R_{\rm crit}$ where $M_{d}/M_{\rm BH}\sim1$, 
i.e.\ where the potential is only transitioning to Keplerian, 
and where it can be globally unstable.
The pattern speed $\Omega_{p}$ will simply reflect $\Omega(R_{\rm crit})$. 
But f the mode can propagate inwards at constant $\Omega_{p}$, 
it will eventually be a slow mode, relative to the local $\Omega$.
This is, in fact, what is seen in simulations 
(see \S~\ref{sec:sims} below).

\subsection{Mode Structure and Propagation}
\label{sec:structure}

How does this occur? For now, we will remain in the WKB approximation and 
consider how such a mode (stable or unstable) might evolve. 
Given a mode, the wave packets themselves propagate with 
approximate group velocity  
$v_{g}={\rm d}\omega/{\rm d}k = {\rm sign}(k)\,(c_{s}^{2}-G\,\Sigma)/(\omega-\Omega)$ or 
$\pi\,G\,\Sigma\,\Omega^{-1}+2\,c_{s}^{2}\,k\,\Omega^{-1}$ for 
slow modes. For a cold disk this is simply $v_{g}\approx(\omega-\varomega)\,R\,|kR|^{-1}$; 
and since $\omega\sim\Omega(R_{\rm crit})$ and the mode is global, 
this is $\sim V_{c}(R_{\rm crit})$. The timescale for the mode to travel is just the 
dynamical time at this critical radius. 

If the mass profile is too shallow, and $c_{s}$ or $\sigma$ remains constant at small 
radii, then the wave will refract back at some $Q$ barrier at some minimum 
radius (for constant $\beta$, refraction occurs with $\eta<1/2$, the 
same criteria that \citet{ostriker:eccentric.waves.via.forcing} show applies for 
modes in a pure fluid disk with a hard outer edge). In non-linear simulations, 
this typically leads to pile-up of inflows, gradually steepening the profile; 
the consequences of this for setting galaxy profile shapes is 
discussed in \citet{hopkins:cusp.slopes}. Here, the mass profile is fixed; 
%\footnote{\citet{ostriker:eccentric.waves.via.forcing} 
%show that the same restriction applies for modes in a pure fluid disk with a hard outer edge.}
%Provided the gas supply is sufficient, 
%this will ``pile up'' inflows near this refraction radius until the profile is steepened to allow inflows. 
but provided the mass profile is sufficiently steep such that the RHS of 
Equation~\ref{eqn:slowmode.dispersion.gas} 
remains finite as $r\rightarrow0$, then modes can propagate through to $R=0$. 

Provided that the sound speed is finite, wave packets in a gaseous disk 
can propagate through the OLR to $r\rightarrow\infty$, eventually becoming 
simple sound waves. This is discussed in 
\citet{adams89:eccentric.instab.in.keplerian.disks} -- because the waves 
can freely escape carrying the mode energy and angular momentum (and will 
reflect off small radii as above), 
infinite pure gaseous disks in nearly-Keplerian potentials do not support strong 
growing modes. Instead, for gaseous disks, mode growth is sensitive to the 
description of the disk edge, and if a ``hard'' edge is assumed, specifically requires efficient reflection of waves off the outer 
edge. As a consequence, it is difficult to determine the growth rates for a pure gaseous disk 
without some {\em a priori} knowledge of the edge structure, and the derivations therein cannot 
be generalized in a straightforward manner to disks with smooth (or thick) edges. 
However, for a stellar disk, the mode cannot propagate beyond the 
OLR where $\Delta\equiv\kappa^{2}-m\,(\Omega-\Omega_{p})^{2}=0$ (this 
acts as an effective outer edge). 
Refraction of the stellar waves off this boundary is important, and means that 
mode growth is possible even when the disk extends to $R\gg R_{\rm OLR}$. 
Together, these constraints set the dynamic range of the mode. 

Physically, how can the mode propagate, if it ``begins'' at larger radii?
It is easy to see as 
the eccentric mode at outer radii exciting 
strong eccentric perturbations at smaller radii. 
For a slow $m=1$ mode near the BH (where the near-Keplerian approximation is good), 
the equations of motion for the perturbed 
velocity 
${\bf v} = R\,\Omega\,\hat{\phi} + v_{r}\,\hat{R} + v_{\phi}\,\hat{\phi}$ 
become, at this order, 
\begin{align}
\label{eqn:vrperturb}
v_{r} &= -\frac{i}{2\,(\omega_{p}-\varomega)}\,\left(  
\frac{{d}\Phi_{e}}{{d}r} + \frac{2\,\Phi_{e}}{r}\right)  \\ 
v_{\phi} &=\frac{i}{2}\,v_{r}
\end{align}
where $\Phi_{e}$ is the ``external'' perturbing potential from the 
larger-scale mode. 

Consider an annulus $R_{1}$, down to which the mode has 
efficiently propagated, and a slightly interior radius $R_{0}$, 
which remains unperturbed. In the WKB limit the 
perturbing potential is dominated by the local structure -- 
so just interior to $R_{1}$ it is 
$\approx \Phi_{1}(R_{1})=2\pi\,G\,\Sigma_{1}(R_{1})\,|k|^{-1}$. 
There is no local corrugation present at $R_{0}$ or interior (by 
definition) to cancel this perturbation term. 
This applies as well in the global (non-WKB) limit; 
consider the limiting case of 
an element on a circular orbit inside an eccentric ring with mass 
$M_{\rm ring}$ at radius $R_{1}$ (with $m=1$ mode amplitude 
$|a|$ defining the eccentricity of the ring). The magnitude of the 
asymmetric term in the potential just inside $R_{1}$ is 
$\approx |a|\,G\,M_{\rm ring}/R_{1}\sim \pi\,G\,\Sigma_{1}(R_{1})\,R_{1}$. 
If the disk is cold, then the local pattern speed 
is given directly by $\omega-\varomega\approx\pi\,G\,\Sigma_{0}\,|k|\,\Omega^{-1}$. 
Taking these estimates for the potential perturbation and 
pattern speed and applying them in Equation~\ref{eqn:vrperturb}, we obtain 
\begin{align}
v_{r} &= -\frac{\Sigma_{1}}{\Sigma_{0}}\,|kR|^{-1}\,\Omega\,R\  \\
|{\bf e}| &\sim \left| \frac{v_{r}}{V_{c}} \right| = \frac{\Sigma_{1}}{\Sigma_{0}}\,|kR|^{-1}= \frac{|a|}{|kR|}\sim|a|\ .
\end{align} 
For non-trivial mode amplitude 
$|a|\sim\Sigma_{1}/\Sigma_{0}$, and a global mode $|kR|\sim 1$, 
corresponding eccentricities and coherent $m=1$ mode amplitudes are induced. 
This of course can then induce eccentricity at the next smaller annulus, and so on, 
allowing the perturbation to grow even at small $R$. 
By the same arguments, at much steeper slopes $\eta\gtrsim1$, 
the system will become more ``stiff'' against inwards propagation of 
eccentricity, but there is no strict cutoff/refraction 
\citep[see also][who find the same for disks of 
planets and planetesimals]{zakamska:eccentricity.wave.propagation}. 

The key facet to note is that for an ``external'' driver of the perturbation $\Phi_{e}$, 
the response is large independent of the local radio of disk to BH mass, 
and is linear in $\Phi_{e}$. 
Thus, the mode only needs to be locally unstable {\em somewhere} in order to 
self-amplify (recall, our stability analysis is in the WKB limit and hence local). 
Self-gravity will grow the strength of the mode there, but the system will respond 
coherently at small radii, even where the system is nominally stable (a local mode 
there would not self-amplify).

Finally, in both gas and stars, near the inner refraction radius (for $\eta\lesssim1/2$), 
an initially global mode must wind up to $|k|\sim\Omega\,c_{s}^{-1}$ or 
$|k|\sim \Omega\,\sigma_{z}^{-1}$, respectively. 
Near the OLR or $r\rightarrow0$ (if $\eta\gtrsim1/2$) in stars, then 
$|kR|\,\exp{\{  -\beta\,|kR|  \}}\rightarrow0$ so there are the 
standard two branches: long $|kR|< \beta$ 
and short $|kR| > \beta$. The long branch solution corresponds to the 
extension of the $g$-modes described in \citet{tremaine:slow.keplerian.modes}, 
but for $\eta\gtrsim1/2$ and an OLR out at large radii where $M_{d}/M_{\rm BH}$ 
is not small, they do not have to have negative (retrograde) pattern speeds. 
The short branch corresponds to the $p$-modes described there.

\begin{figure*}
    \centering
    \scaleup
    \plotside{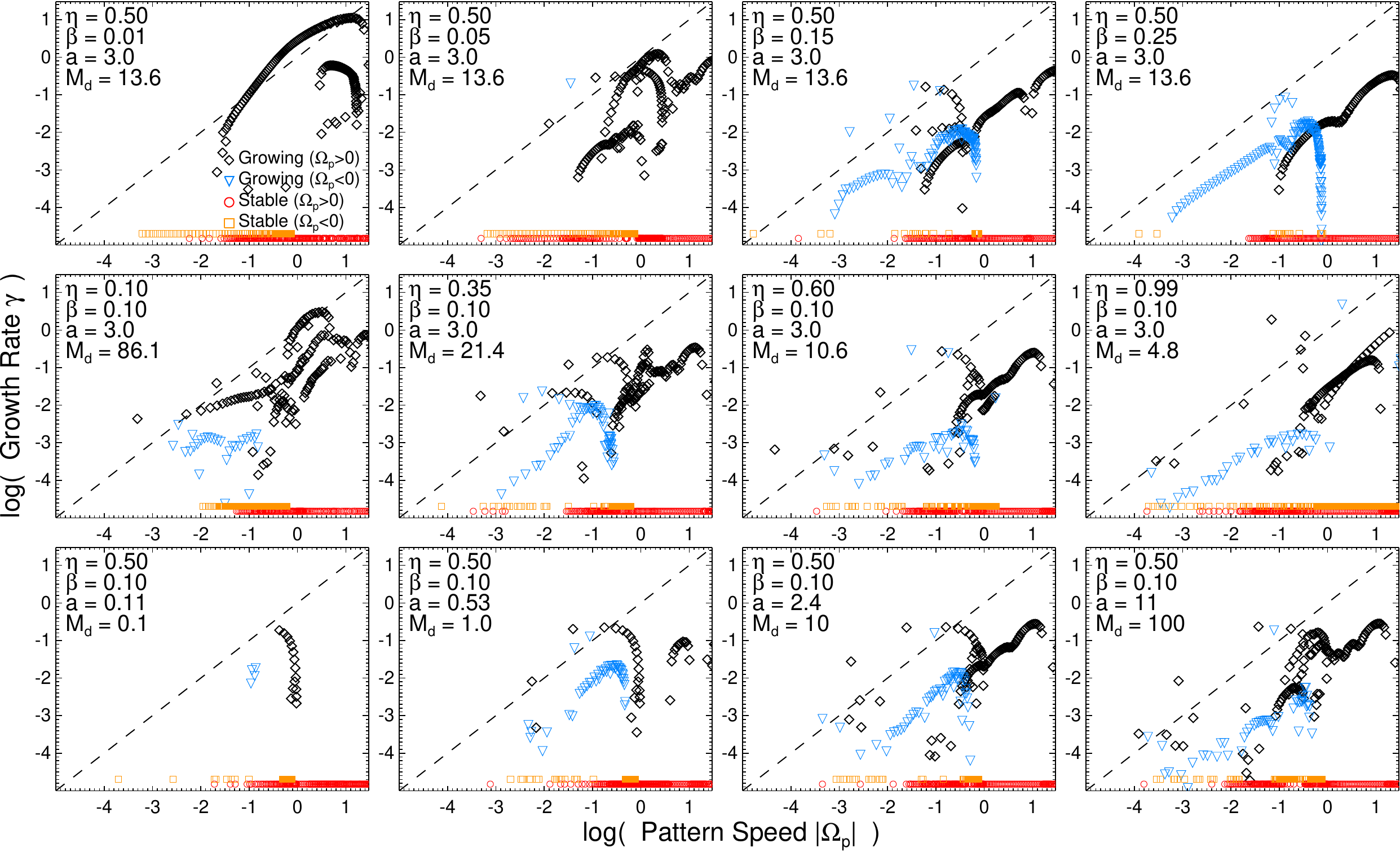}
    %\plotone{f1.pdf}
    \caption{Spectrum of eigenvalues (normal mode frequencies) for $m=1$ modes in 
    the truncated power-law disk in Equation~\ref{eqn:sigma.cutoff} around a BH, 
    in units of $M_{\rm BH}=R_{0}=G=1$, as a function of the parameters 
    $\beta$ (the disk softening $\sim h/R$), $\eta$ (the mass profile slope) 
    and $a$ (the outer disk scale length, which determines the disk-to-BH mass ratio 
    $M_{d}/M_{\rm BH}$). Black diamonds show the pattern speed $\Omega_{p}$ and growth 
    rate $\gamma$ for growing modes ($\gamma>0$). 
    Stable modes ($\gamma=0$) are shown for comparison (red circles at $\gamma=10^{-5}$ 
    just to be displayed). Blue triangles show growing 
    modes with $\Omega_{p}<0$ (retrograde precession), plotting $|\Omega_{p}|$. 
    For comparison, the dashed line corresponds to $\gamma=|\Omega_{p}|$. 
    There is a large spectrum of (mostly prograde) 
    rapidly-growing modes with $\gamma\sim\Omega_{p}$, 
    and growth rates that increase in colder and relatively more massive disks. 
    Non-zero growth rates can be present even in disks that are locally stable 
    ($Q\gg1$) and disks with small $M_{d}/M_{\rm BH} \sim 0.1$. 
    \label{fig:eigenvalues}}
\end{figure*}

\begin{figure}
    \centering
    \scaleup
    \plotone{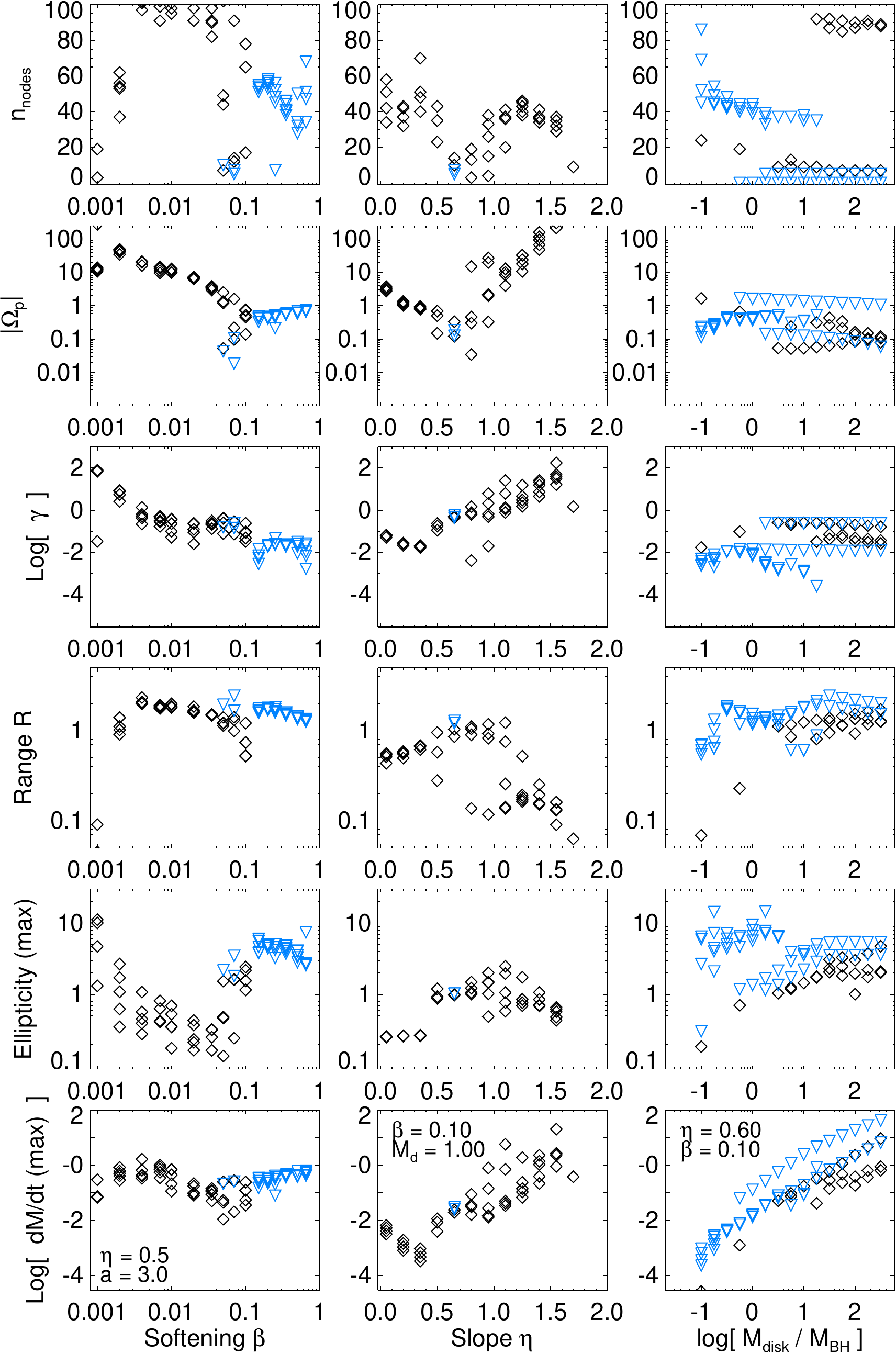}
    %\plotone{f1.pdf}
    \caption{Scaling of the largest dynamic-range growing 
    ($\gamma>0$) normal modes 
    (largest five $\mathcal{R}\equiv\int |a|\,{d\log R}$). 
    {\em Left:} At fixed slope and $a$, $M_{d}$, as a function of 
    $\beta$. {\em Center:} As a function of slope $\eta$ 
    (varying $a$ to hold $M_{d}=M_{\rm BH}$). 
    {\em Right:} As a function of $M_{d}$ (and $a$). 
    In each column, we show the number of 
    zero-crossings $n$, the mode pattern speed 
    $|\Omega_{p}|$ (black have 
    $\Omega_{p}>0$, blue have $\Omega_{p}<0$), 
    the growth rate $\gamma$, the 
    range $\mathcal{R}$, the maximum ellipticity 
    $\equiv R_{1}/R$ induced by the mode (mode amplitude 
    normalized so that ${\rm MAX}(|a|)=1$), 
    and the maximum inflow rate induced 
    (units of $M_{0}\,(G\,M_{0}/R_{0}^{3})^{-3/2}$; 
    for a BH this is 
    $8400\,f_{\rm gas}\,\msun\,{\rm yr^{-1}}\,(M_{\rm BH}/10^{8}\,\msun)^{3/2}\,(R_{0}/4\,{\rm pc})^{-3/2}$; 
    for a protostellar/circumstellar disk 
    $0.2\,f_{\rm gas}\,\msun\,{\rm yr^{-1}}\,(M_{\ast}/\msun)^{3/2}\,(R_{0}/10\,{\rm au})^{-3/2}$). 
    \label{fig:eigenvalue.scalings}}
\end{figure}

\begin{figure}
    \centering
    \scaleup
    \plotone{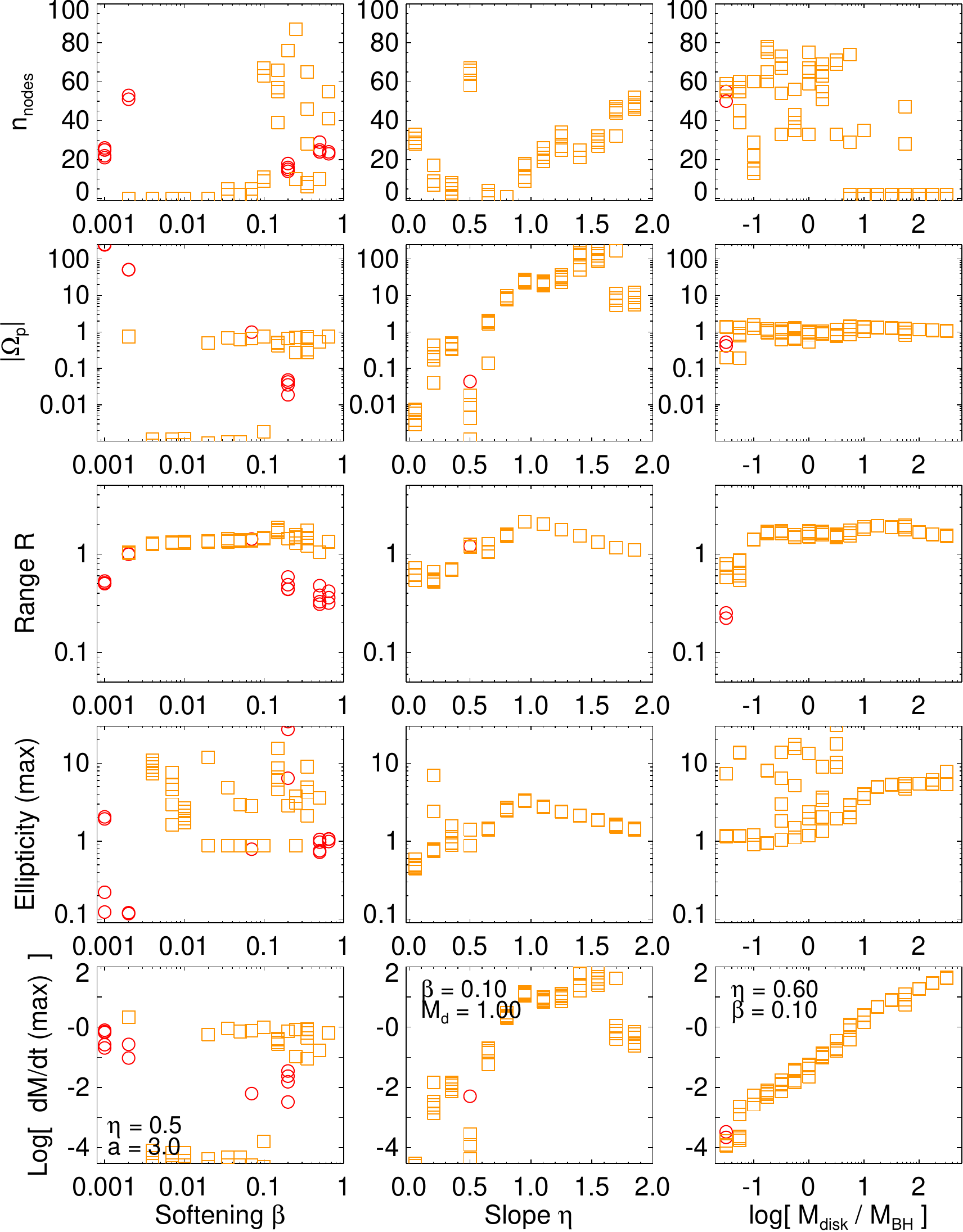}
    %\plotone{f1.pdf}
    \caption{As Figure~\ref{fig:eigenvalue.scalings}, but for 
    the largest dynamic range stable ($\gamma=0$) modes. 
    The scalings are similar -- the major difference is the existence of stable 
    (but not unstable) modes at $M_{d}\ll M_{\rm BH}$. 
    \label{fig:eigenvalue.scalings.stable}}
\end{figure}

\section{Exact Solutions}
\label{sec:exact}

\begin{figure*}
    \centering
    \scaleup
    \plotside{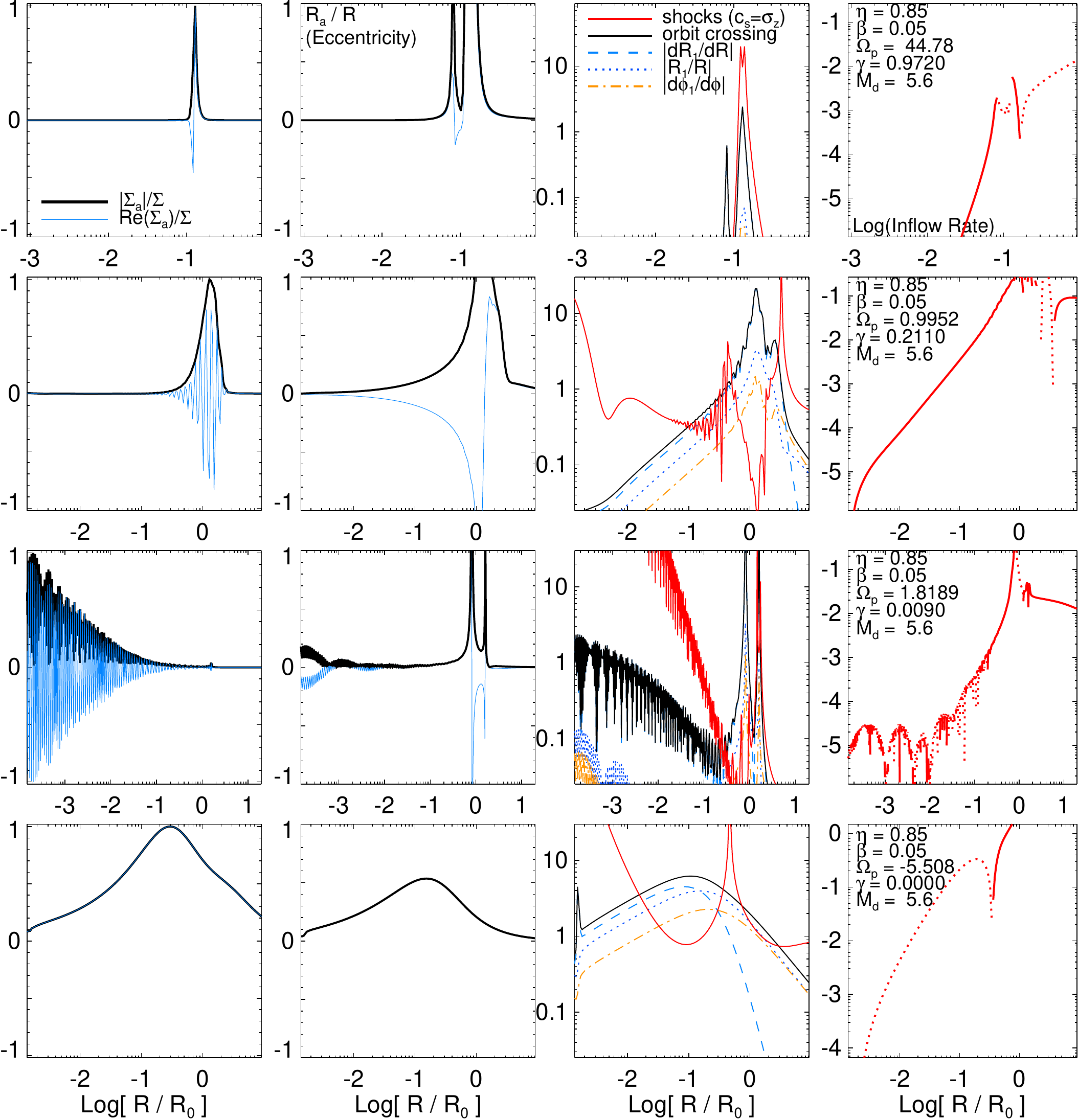}
    %\plotone{f1.pdf}
    \caption{Example eigenmodes of the $m=1$ mode in a 
    BH+power-law disk with softened gravity. 
    Each row shows the properties of one such mode, with 
    parameters at right (disk mass profile slope $\Sigma\propto R^{-\eta}$, 
    softening $\beta\sim h/R$, pattern speed $\Omega_{p}={\rm Re}(\omega)$, 
    mode growth rate $\gamma={\rm Im}(\omega)$, and disk-to-BH mass ratio $M_{\rm disk}$). 
    Radii are in units of $R_{0}$ (the radius where the enclosed disk mass 
    $\sim M_{\rm BH}$). Modes are shown from an outer radius outside the OLR 
    (where they terminate) to $\sim10^{-3}\,R_{\rm OLR}$. 
    {\em Left:} Mode amplitude. 
    The absolute magnitude of the density perturbation 
    $|\Sigma_{a}|/\Sigma$, and ${\rm Re}(\Sigma_{a}/\Sigma)$ (the latter shows 
    the wavenumber, i.e.\ $\exp{[i\,\int\,k\,dr]}$). 
    {\em Center-Left:} Induced radial perturbation $\equiv R_{a}/R$ 
    (equal to the eccentricity, in the near-Keplerian regime). 
    {\em Center-Right:} Conditions for orbit-crossing. 
    The perturbation radii $|R_{1}/R|$, $\zeta=|dR_{1}/dR|$, 
    and $|d\phi_{1}/d\phi|$. The ``all'' line is 
    $(|R_{1}/R|^{2}+|dR_{1}/dR|^{2}+|d\phi_{1}/d\phi|^{2})^{1/2}$. 
    If any are $>1$, there are orbit crossings. 
    Red line shows the criterion from Equation~\ref{eqn:shocks}, 
    for where gas shocks will be induced (when $>1$); note this can occur even without formal orbit 
    crossings. 
    {\em Right:} Log of the gas
    inflow (solid) or outflow (dotted) rate $|\dot{M}|$ driven by the mode, 
    in units of $f_{\rm gas}\,M_{\rm BH}\,\Omega_{0}$, where 
    $\Omega_{0}\equiv(G\,M_{\rm BH}\,R_{0}^{-3})^{1/2}$. 
    In the units here, $1$ corresponds to 
    $8400\,\msun\,yr^{-1}\,(M_{\rm BH}/10^{8}\,\msun)^{3/2}\,(R_{0}/4\,{\rm pc})^{-3/2}$ 
    for a BH, or 
    $0.20\,\msun\,{\rm yr^{-1}}\,(M_{\ast}/\msun)^{3/2}\,(R_{0}/10\,{\rm au})^{-3/2}$ for a 
    circumstellar disk. 
    The (arbitrary) mode amplitudes are normalized so ${\rm MAX}(|\Sigma_{a}|/\Sigma)=1$.  
    The figure shows several modes with different $\Omega_{p}$ and $\gamma$ in a 
    fixed system. The largest $\gamma$ and $\Omega_{p}$ mode ({\em top}) is local; 
    at lower $\Omega_{p}$ ({\em middle}) the modes cover a larger dynamic range, but still 
    have moderate $|kR|$; the most global mode ({\em bottom}) is stable, and retrograde. 
    \label{fig:m1.1}}
\end{figure*}
\begin{figure}
    \centering
    \scaleup
    \plotone{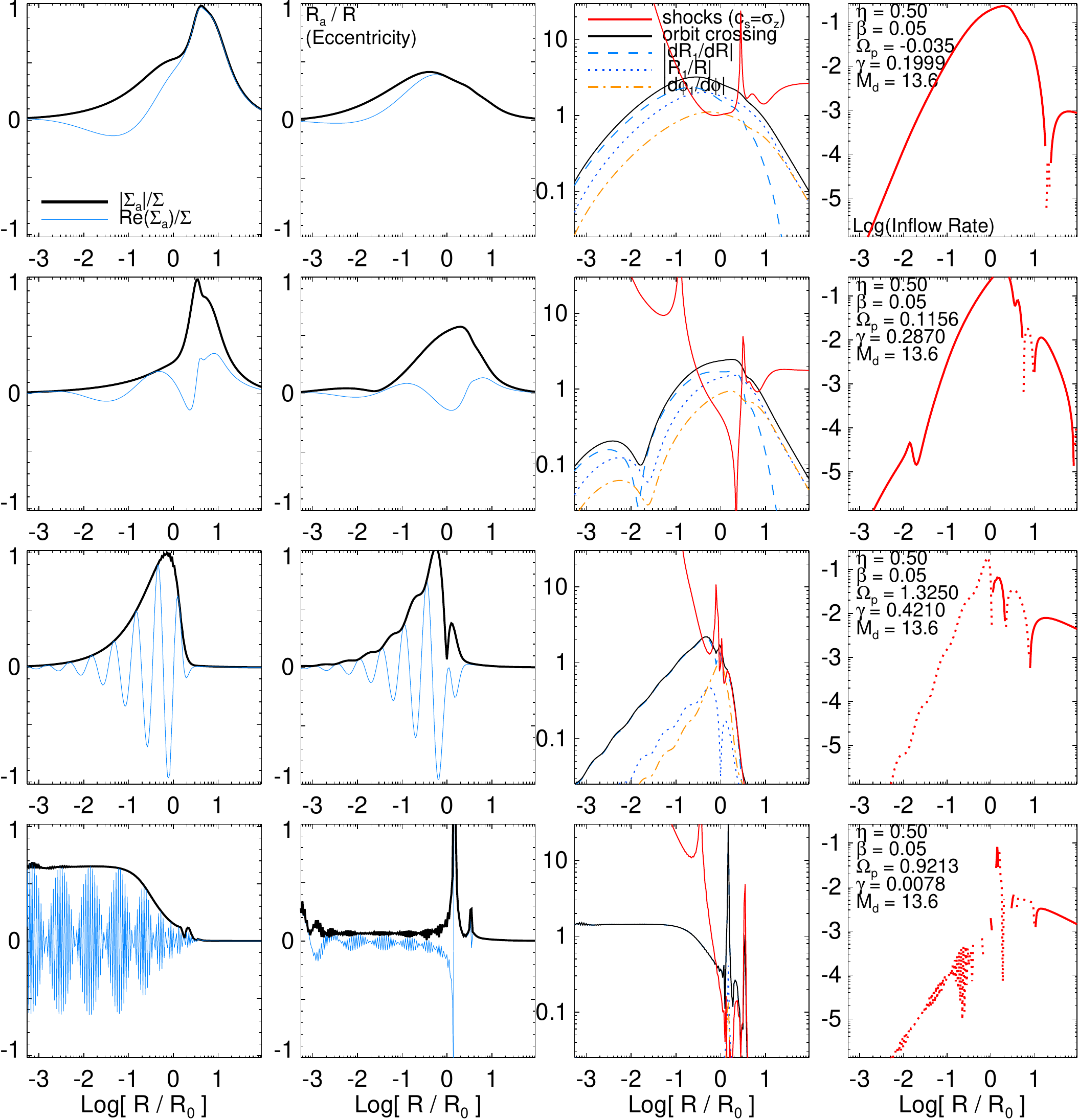}
    %\plotone{f1.pdf}
    \caption{Figure~\ref{fig:m1.1} continued. Mode structure versus $n$, the number of nodes, 
    in the growing modes of an otherwise fixed system. 
    From top to bottom, $n$ increases from $2$ to $>100$. The 
    moderate-$n$ modes have similar $\Omega_{p}$ 
    and larger $\gamma$. Despite large $n$, some of these modes are clearly very long-wavelength. 
    \label{fig:m1.2}}
\end{figure}
\begin{figure}
    \centering
    \scaleup
    \plotone{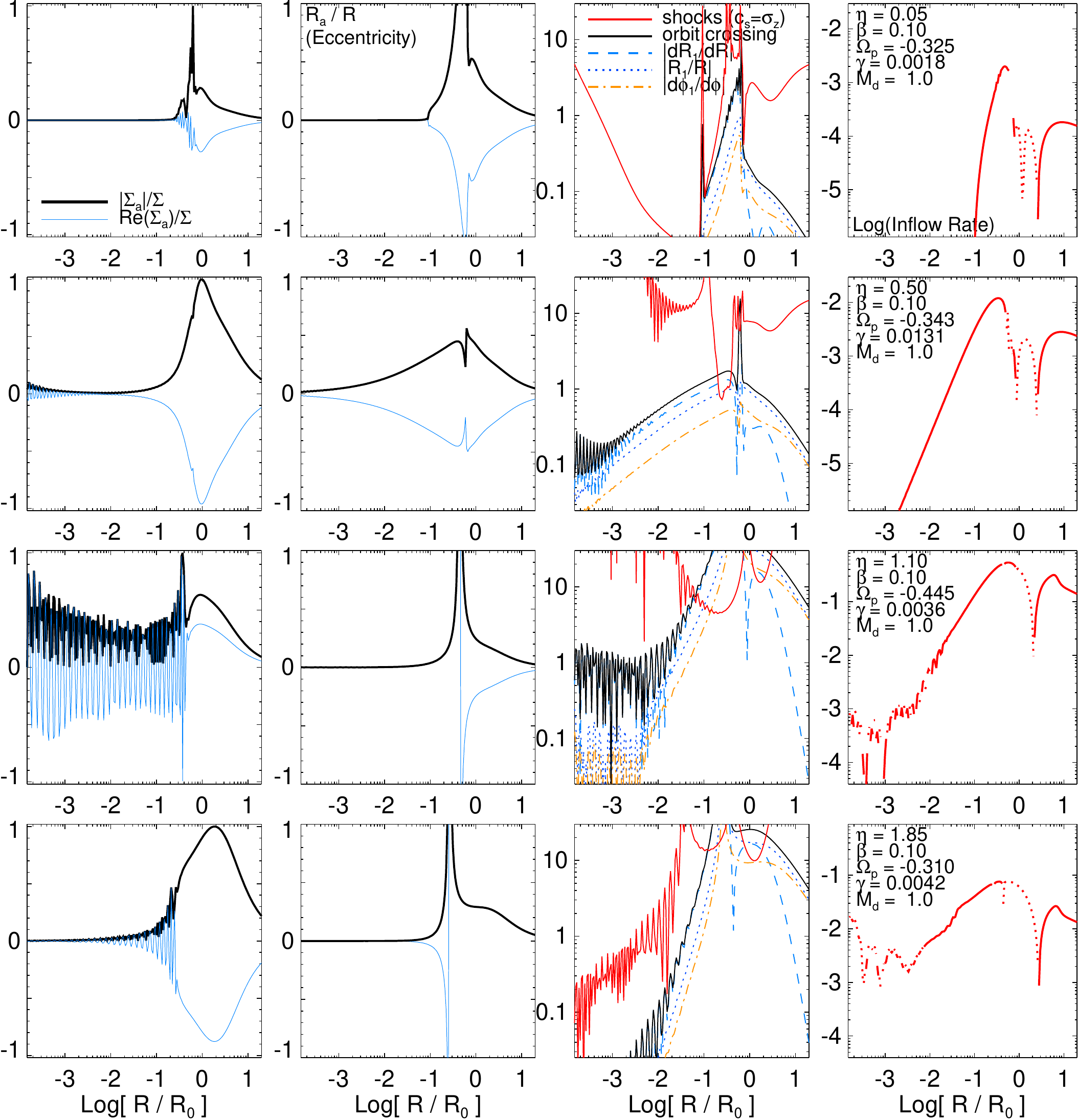}
    %\plotone{f1.pdf}
    \caption{Figure~\ref{fig:m1.1} continued. Mode structure versus mass profile slope 
    $\eta$ at otherwise fixed properties (top-to-bottom: $\eta=0.05,\,0.5,\,1.1,\,1.85$). 
    For shallow slopes $\eta<1/2$, the mode at fixed $\beta$ cannot be supported at 
    $R\rightarrow0$, and so it is localized. 
    At larger $\eta$, the modes carry amplitude to $R\rightarrow0$. 
    %At $\eta >1$, however, the inwards propagation efficiency of eccentricity 
    %decreases (the disk is ``stiffer'' against external perturbation. 
    \label{fig:m1.3}}
\end{figure}
\begin{figure}
    \centering
    \scaleup
    \plotone{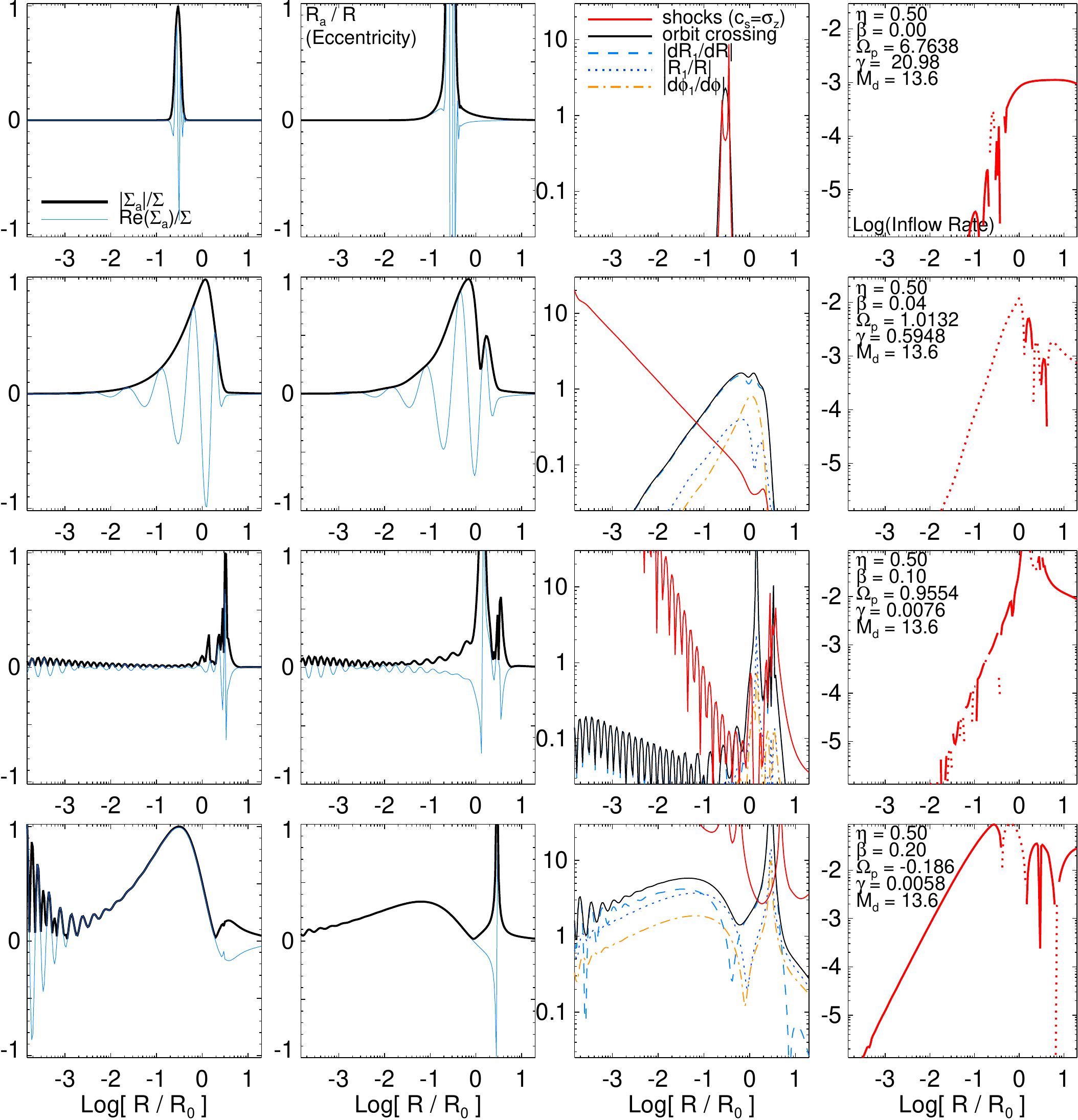}
    %\plotone{f1.pdf}
    \caption{Figure~\ref{fig:m1.1} continued. Mode structure versus gravitational softening/disk 
    scale height at otherwise fixed properties (top-to-bottom: $\beta=0.002,\,0.04,\,0.1,\,0.2$). 
    Modes with otherwise similar properties are more global in softened/thick 
    disks, and more tightly-wound in cold disks. A sufficiently cold disk does not support 
    the same global modes because the implied growth rate at large radii is much larger than 
    that at small radii. 
    \label{fig:m1.4}}
\end{figure}
\begin{figure}
    \centering
    \scaleup
    \plotone{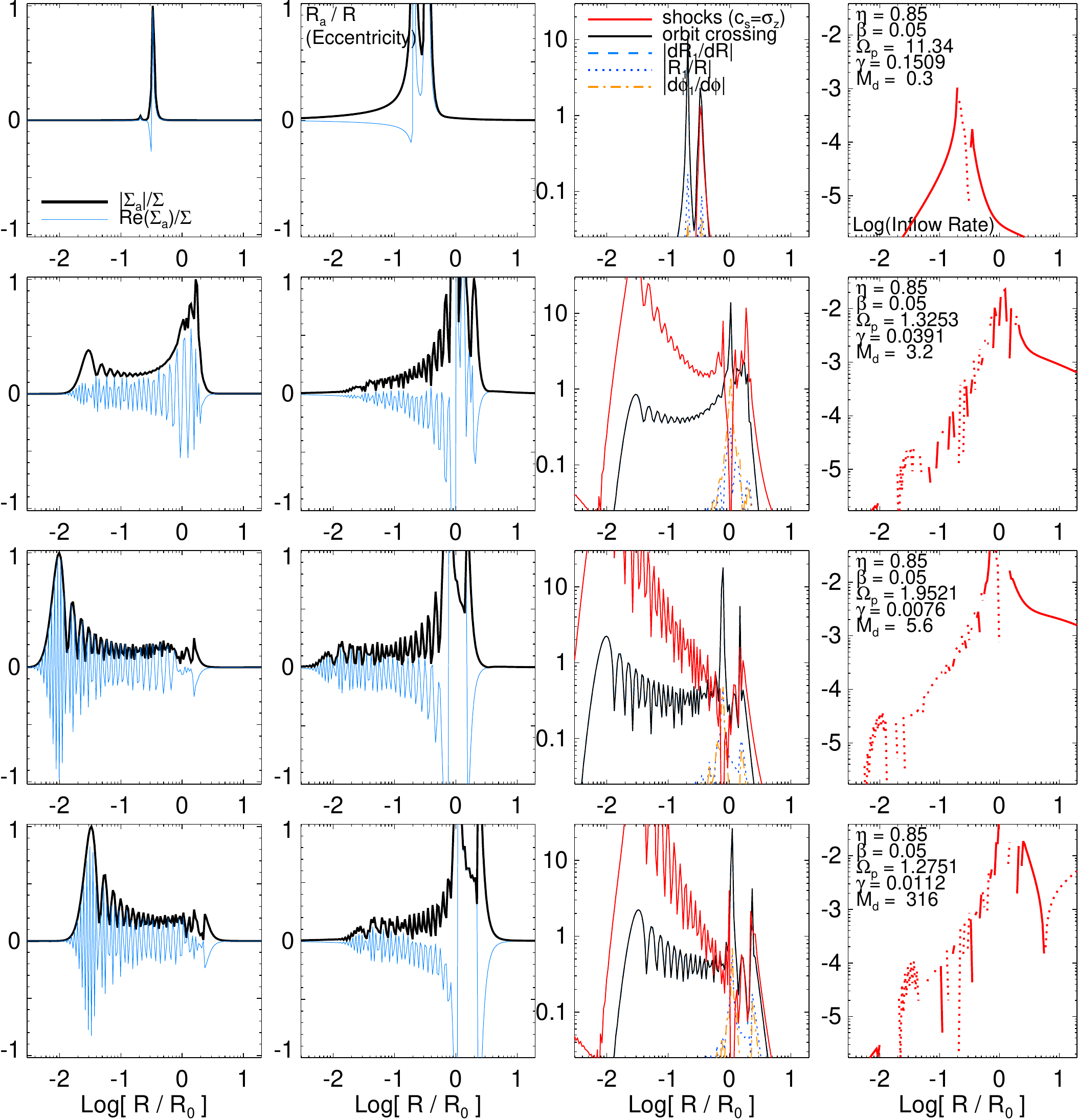}
    %\plotone{f1.pdf}
    \caption{Figure~\ref{fig:m1.1} continued. Mode structure versus 
    disk-to-BH mass ratio at otherwise fixed properties (top-to-bottom: 
    $M_{\rm disk}/M_{\rm BH}=0.31,\,3.2,\,5.6,\,316.3$). 
    At low $M_{\rm disk}$, the most unstable (local, high-$k$) modes appear first. 
    Higher $M_{\rm disk}\sim M_{\rm BH}$ allows more global modes. 
    Once $M_{\rm disk}\gtrsim M_{\rm BH}$, the structure of modes that exist at small 
    radii (the nearly-Keplerian regime) is similar regardless of $M_{\rm disk}$. 
    \label{fig:m1.5}}
\end{figure}

The WKB approximation has been instructive. However, this does not allow 
us to survey the complete parameter space, and it is a suspect approximation 
for any global mode with small $|kR|$. Therefore, it is important to check our 
conclusions and examine the mode structure in exact solutions to the perturbed 
linear equations of motion. 

\subsection{Equations of Motion}

We must choose a specific disk model, so for convenience, adopt the 
power-law model in \S~\ref{sec:definitions}. 
We can freely define $R_{0}$ anywhere; so choose it 
such that (for the infinite disk) $M_{d}(<R_{0})=[\alpha\,(2-\eta)]^{-1}\, M_{\rm BH}$ -- 
i.e.\ $\Sigma_{0}=M_{\rm BH}/(2\pi\,\alpha\,R_{0}^{2})$. 
We can then still apply any disk cutoff with $a$ in units of $R_{0}$ determining 
the ratio $M_{d}/M_{\rm BH}$. 
We will consider all numerical quantities in units $M_{\rm BH}=G=R_{0}=1$. 
%so that 
%\begin{align}
%\Omega^{2} &=\frac{1}{r^{3}}+r^{-(\eta+1)} \\ 
%\kappa^{2} &=\frac{1}{r^{3}} + (3-\eta)\,r^{-(\eta+1)}
%\end{align}

Following \citet{tremaine:slow.keplerian.modes}, define the perturbed quantities 
as $\Sigma(r)+\Sigma_{1}({\bf r},t)$, 
${\bf v}_{d}(r)+{\bf v}_{1}({\bf r},t)=r\,\Omega\,\hat{\phi} + u_{1}({\bf r},t)\,\hat{r} + v_{1}({\bf r},t)\,\hat{\phi}$, 
$\Phi(r)+\Phi_{1}({\bf r},t)$, 
and in standard fashion write the perturbation variables 
in the form $X_{1}({\bf r},t)=X_{1}(r,\phi,t)=X_{a}(r)\,\exp{[i\,(\phi-\omega\,t )]}$. 
The equations of motion, together with Poisson's equation, can be written
\begin{align}
i\,(\Omega-\omega)\,\Sigma_{a} &= -\frac{1}{r}\frac{d}{dr}(r\,\Sigma\,u_{a}) - \frac{i}{r}\,\Sigma\,v_{a} \\ 
u_{a} &= -\frac{i}{\Delta}\left[(\Omega-\omega)\,\frac{d}{dr} + \frac{2\,\Omega}{r}   \right]\,\Phi_{a} \\ 
v_{a} &= \frac{1}{\Delta}\left[\frac{\kappa^{2}}{2\,\Omega}\,\frac{d}{dr} + \frac{(\Omega-\omega)}{r}   
\right]\,\Phi_{a} \\ 
\Delta &= \kappa^{2} - (\Omega-\omega)^{2} \\ 
\Phi_{a} &= \int_{0}^{\infty}\,{dr^{\prime}}\,{r^{\prime}}\,P(r,\,r^{\prime})\,\Sigma_{a}(r^{\prime}) \\ 
P(r,\,r^{\prime}) &= P_{\rm direct}(r,\,r^{\prime}) + P_{\rm indirect}(r,\,r^{\prime}) \\
&= 
%-\frac{\pi\,G}{r_{>}}\,b_{1/2}(r_{<}/r_{>}) + \frac{\pi\,G\,r^{\prime}}{r^{2}}\nonumber
-\frac{\pi\,G}{r_{>}}\,b_{1/2}(r_{<}/r_{>}) + 
\frac{\pi\,G\,r}{r^{\prime\ 2}}\nonumber
%\frac{\pi\,\omega^{2}\,r\,r^{\prime}}{M_{\rm tot}}\nonumber
\end{align}
where the kernel $P$ for the potential includes the direct 
(what would appear even in an inertial frame) 
and indirect components. 
The Laplace coefficient $b_{1/2}$ is given by
\be
b_{1/2}(x)=\frac{2}{\pi}\int_{0}^{\pi}\,\frac{\cos{\theta}\,{d\theta}}
{(1-2\,x\,\cos{\theta}+x^{2}+\beta^{2})^{1/2}}
\ee
where $\beta$ represents the gravitational softening.\footnote{Note that 
when we soften the gravity for the perturbation, we also soften that of the 
unperturbed disk. This leads to a slight change in the disk potential, 
and hence $\Omega$ and $\kappa$ given in \S~\ref{sec:definitions} for 
a cold disk. 
However, the difference is very small (usually $<2\%$), and has no effect on 
our conclusions (compared to simply using $\Omega(R)$ of a cold disk). 
} 
We choose this particular softening both because it is numerically 
convenient, and because it reduces exactly to the solution for a 
disk with scale height $\beta=h/R$ 
(with $\rho=\Sigma/2\,h$ at $|z|<h$) when $h\ll R$. 

This formulation of the indirect potential follows \citet{murraydermott}. 
The inclusion of the indirect potential is necessary because 
the $m=1$ mode changes the center of mass, leading to motion by the 
BH about that center of mass. Recall that our coordinate frame is centered on the BH, 
thus $r$ is the vector distance to the BH, and the frame is rotating 
about the BH. 

%This formulation of the indirect potential follows \citet{adams89:eccentric.instab.in.keplerian.disks}, 
%where $M_{\rm tot}$ is the total 
%system mass $=M_{\rm BH}+M_{\rm disk}(<R_{\rm max})$, 
%$R_{\rm max}$ being the maximum radius of the disk out to which 
%we integrate. The inclusion of the indirect potential is necessary because 
%the $m=1$ mode changes the center of mass, and makes some 
%small difference around $R_{0}$, but in general it has weak effects 
%(at small radii, it declines both $\propto r$ and enters only at second-order 
%in $\omega/\Omega$). And if the outer boundary of the disk is sufficiently 
%large such that $M_{\rm tot}\gg M_{\rm BH}$, its effects vanish.

For a given $\omega$, $\eta$, and $\beta$, 
then, the solution to these coupled equations determines 
the perturbation structure. 
Because we are interested in the behavior where $\omega/\Omega$ 
may be non-trivial, the equations are non-linear in $\omega$. 
However, we can combine the equation in $\Sigma_{a}$, $u_{a}$, and $v_{a}$ to obtain 
a single equation 
\begin{align}
\nonumber 0 = & -(\Omega-\omega)\,\Sigma_{a}  \\ 
\nonumber & 
+\frac{\Sigma}{r^{2}\,\Delta}\,\left[ 2\,\Omega\,(\nu_{\Sigma}+\nu_{\Omega}-\nu_{\Delta}) - (\Omega-\omega)
\right]\,\Phi_{a}  \\ 
\nonumber & 
+\frac{\Sigma}{r\,\Delta}\,\left[ (\Omega-\omega)\,(1+\nu_{\Sigma}-\nu_{\Delta}) + \Omega\,(2+\nu_{\Omega}) - 
\frac{\kappa^{2}}{2\,\Omega} \right]\,\Phi_{a}^{\prime}  \\ 
& 
+\frac{\Sigma}{\Delta}\,\left[ \Omega-\omega\right]\,\Phi_{a}^{\prime\prime}
\label{eqn:mode.eom}
\end{align}
Where $\Phi^{\prime} = \int\,{dr^{\prime}}\,r^{\prime}\,(\partial P(r,\,r^{\prime})/\partial r)\,\Sigma_{a}(r^{\prime})$ 
and $\Phi^{\prime\prime} = 
\int\,{dr^{\prime}}\,r^{\prime}\,(\partial^{2} P(r,\,r^{\prime})/\partial r^{2})\,\Sigma_{a}(r^{\prime})$, 
and $\nu_{X}\equiv \partial \ln{X}/\partial \ln{r}$. 
If we discretize $\Sigma_{a}$ into some grid in $\log{r}$ and apply some summation rule to 
evaluate the integrals in $\Phi$, then we can write this as a linear operator acting on the 
perturbed density, 
\be 
{\bf M}\,\bar{\Sigma}_{a}(\bar{r}^{\prime}) = \bar{0}
\ee

The eigenvalues $\omega$ and eigenvectors $\bar{\Sigma}_{a}$ represent the exact 
homogenous solutions to this equation. 
In practice, we obtain solutions following the method outlined in Appendix~B of 
\citet{adams89:eccentric.instab.in.keplerian.disks}. 
If we multiply Equation~\ref{eqn:mode.eom} by 
$\Delta^{2}$, then we can eliminate all occurrences of $\omega$ in the 
denominator and write the resulting matrix equation as 
a fifth-order equation in $\omega$: 
$({\bf M}_{0}+\omega\,{\bf M}_{1}+\omega^{2}\,{\bf M}_{2}
+\omega^{3}\,{\bf M}_{3}+\omega^{4}\,{\bf M}_{4}+\omega^{5}\,{\bf M}_{5})\,
\bar{\Sigma}_{a}(\bar{r}^{\prime})=0$ 
where the ${\bf M}_{n}$ are independent of $\omega$. 
With the appropriate substitutions, this can then be turned into a 
single eigenvalue equation ${\bf M}_{\rm 5x5}\,\bar{T}_{a} = \omega\,\bar{T}_{a}$, 
where if ${\bf M}$ is $NxN$ elements, ${\bf M}_{\rm 5x5}$ is $5Nx5N$, 
and $\bar{T}_{a} = 
(\bar{\Sigma}_{a},\,
\mathcal{O}(\omega\,{\bf M}_{n})^{1}\,\bar{\Sigma}_{a},\,
\mathcal{O}(\omega\,{\bf M}_{n})^{2}\,\bar{\Sigma}_{a},\,
\mathcal{O}(\omega\,{\bf M}_{n})^{3}\,\bar{\Sigma}_{a},\,
\mathcal{O}(\omega\,{\bf M}_{n})^{4}\,\bar{\Sigma}_{a})$
is constructed from combinations of the ${\bf M}_{n}$, $\bar{\Sigma}_{a}$ and $\omega$. 
It is then straightforward to solve for all eigenvalues and eigenvectors. 
%In practice, 
%starting from a given guess at a complex eigenvalue $\omega$, we 
%numerically obtain the 
%exact eigenvalue such that $|| {\bf M} ||=0$, then find the roots of the above 
%matrix equation.
\footnote{Numerically, we typically realize this on a grid of 
$\sim400-4000$ elements evenly spaced in $\log{r}$ from a factor of several outside of the 
OLR, where the mode amplitude vanishes, to $r\sim 10^{-5}\,R_{0}$. 
The results are numerically converged. 
Normalizing such that $| \bar{\Sigma}_{a} | = N$, where 
$N$ is the number of grid points, and $|{\rm Row}({\bf M})|=N$ for each 
row of ${\bf M}$, we obtain solutions 
of ${\bf M}\,\bar{\Sigma}_{a}(\bar{r}) = \bar{\delta}$ with typical 
$|\bar{\delta}|^{2}/N \lesssim 10^{-16}$.} 
%We can also multiply both sides of Equation~\ref{eqn:} by $\Delta^{2}$ and 
%obtain a fifth-order polynomial matrix equation in $\omega$, which can be 
%reduced to a linear eigenvalue problem for a $5N\times5N$ matrix 
%following \citet[][see their appendix ]{adams:}; this is useful, but the 
%expansion introduces a number of spurious apparent eigenvalues and is 
%numerically less robust, so we 
%follow up each by checking $|| {\bf M} ||=0$. There are also a large number 
%of singular modes, which are localized to single grid points and are not 
%interesting. 
%In practice, especially for cold disks, the spectrum of $\omega$ is dense so long 
%as it is in a broadly allowed regime (see e.g.\ the solutions in the nearly-Keplerian 
%limit in \citealt{tremaine:slow.keplerian.modes}, for which the spectrum is 
%nearly continuous). Thus for essentially every $\omega$ inside a certain physically 
%allowed regime, some corresponding approximate solution exists.

\subsection{Results}

\subsubsection{Growth Rates}

Figure~\ref{fig:eigenvalues} illustrates the spectrum of 
eigenvalues of growing modes, in power-law disks, 
as a function of the 
softening ($\beta$), slope ($\eta$), and disk cutoff radius/mass 
($a$). 
For now, we simply show all eigenvalues of the above equations 
that fall in the plotted range, making no discrimination 
between local or global modes. We focus on the growing modes, but see that in all 
cases there is a very large spectrum of stable modes; 
there are also decaying modes, but these  
correspond to the complex conjugate pairs of the growing modes shown. 

The nearly continuous lines traced out by the 
eigenvalues correspond to solutions that cover some finite dynamic range 
well inside the cutoff $a$ (i.e.\ where the disk is still scale-free), so that the solution can simply be shifted in 
$r$ and $\omega$. Broadly speaking, the growing modes range in 
growth rates from $\gamma\sim0.01-1\,|\Omega_{p}|$, and the pattern speeds 
$\Omega_{p}$ for both stable and growing modes tend to fall in a similar range. 
We will show how this relates to the characteristic scales of the modes below. 

At fixed $\eta$ and $a$ (or $M_{d}/M_{\rm BH}$), increasing the softening $\beta$ 
decreases the mode growth rates $\gamma/\Omega_{p}$, and decreases the overall 
fraction of unstable modes, as expected. However, growing modes exist even for 
large $\beta\gtrsim0.25$. These are global modes, 
where there is a large contribution to the instability from the indirect potential, 
which is not treated in the WKB limit. Hence, our previous conclusions regarding sufficient 
$\beta$ for stability should be regarded as only pertaining to local instabilities. 
\citet{adams89:eccentric.instab.in.keplerian.disks} found similar results, in that global 
instability could appear even in $Q\gg1$ (everywhere locally stable) disks. 

Modes with moderate or large negative (retrograde) pattern speeds, 
$\Omega_{p}\ll-1$ are generally not supported. 
Some modes with small retrograde pattern speeds are supported, 
increasingly at large $\beta$ and/or $\eta$.
However, such modes will tend to drive outflows (though not 
in every case), so in practice 
may not be able to self-consistently build up 
a disk to the steep surface density profiles ($\eta>1/2$) needed 
to support them in the first place. 

At fixed $\beta$ and $a/R_{0}$, modes with lower $\eta$ (shallower disk profiles) 
have somewhat larger growth rates, but the effect is most pronounced only 
as $\eta\rightarrow0$. This is partly an artifact of holding $a/R_{0}$ 
fixed, since doing so while decreasing $\eta$ means that $M_{d}/M_{\rm BH}$ 
is larger in the lower-$\eta$ disks. But there is also a (small) real effect because 
relatively less of the disk mass is 
at very small radii where the BH dominates the potential. 
The range of unstable modes is similar, though, 
and we see below that these modes propagate over a smaller dynamic range in 
radius. 

At fixed $\eta$ and $\beta$, changing $a$ and correspondingly $M_{d}$ has the expected 
effect -- at larger $M_{d}/M_{\rm BH}$, growth rates (and the fraction of unstable modes) 
increases. Also, the maximum speed of the unstable modes increases, as we might expect 
from the WKB analysis (since the supported mode speeds scale with $M_{d}/M_{\rm BH}$ 
in the nearly-Keplerian limit). There are still significant unstable modes even at low 
$a$ and $M_{d}$ though -- they appear at approximately $M_{d}\sim0.1$. 
So although it is true that in the limit of vanishingly small $M_{d}/M_{\rm BH}$, 
all modes are stable, in practice given the right mass profile and softening, a 
small finite $M_{d}/M_{\rm BH}$ can still give interesting mode growth.

\subsubsection{General Mode Properties}

To see how the mode properties scale in detail with these choices, 
we need to select some subset of modes to study. 
Typically, modes are compared at fixed $n_{\rm nodes}$, the number of 
nodes or zero-crossings. However, because of the large scale-free range, 
there are many modes with the same $n$; moreover, with the large dynamic 
range involved, the $n=0$ modes are not necessarily the ``most global'' 
(they might cover only a limited range in radius). 
Since we are interested in 
modes that cover a large dynamic range, we define the proxy 
\be
\mathcal{R} \equiv \int_{0}^{\infty}\,|a(R)|\,d\log{R}
\ee
where the (arbitrary) normalization of the mode amplitude is set such that 
${\rm MAX}(|a(R)|)=1$ (the maximum value for which the solution would not 
imply negative densities somewhere). 
Roughly speaking, a couple times $\mathcal{R}$ gives the 
number of dex over which the mode has a significant amplitude. 

In Figure~\ref{fig:eigenvalue.scalings}, we select the five growing ($\gamma>0$) 
modes with the largest values of $\mathcal{R}$, and show how their properties 
scale with $\beta$, $\eta$ (in this case, allowing $a$ to vary 
to hold $M_{d}/M_{\rm BH}$ fixed while $\eta$ varies), and the disk mass $M_{d}$. 
In Figure~\ref{fig:eigenvalue.scalings.stable}, we show the same, but for the 
largest-$\mathcal{R}$ stable ($\gamma=0$) modes. 
We show a number of mode properties: the number of nodes 
$n_{\rm nodes}$, the pattern speed $|\Omega_{p}|$, growth rate $\gamma$ 
(for the unstable modes), and 
range $\mathcal{R}$ defined above. We are also interested in the 
effect of the perturbation on eccentricities and, ultimately, inflow. 
The magnitude of the radial perturbation $|R_{1}|/R$ from a given mode is, 
for linear perturbations, 
\be
R_{1} = -\frac{1}{\Delta}\,\left( \frac{d\,\Phi_{1}}{d\,R} + 
\frac{2\,m\,\Omega}{R\,(m\,\Omega-\omega)}\,\Phi_{1} \right)
\ee
Defining the eccentricity as $|{\bf e}|=(r_{a}-r_{p})/(r_{a}+r_{p}) = |R_{1}|/R$, 
we plot the maximum $|{\bf e}|$ induced by the mode over all radii, given 
the same normalization condition ${\rm MAX}(|a(R)|)=1$. 
If the eccentricities are anywhere significant, there can be orbit crossings, 
hence shocks and dissipation. So we can calculate the induced gas inflow 
rates, using the scalings 
derived in \citet{hopkins:inflow.analytics}, for the inflow rates of 
gas driven by shocks induced by gravitational instabilities. 
They show 
\be
\dot{M} = \Sigma_{\rm gas}\,R^{2}\,\Omega\,
{\Bigl |} \frac{\Phi_{1}}{V_{c}^{2}} {\Bigr |}\,
\frac{m\,{\rm sign}(\Omega-\Omega_{p})}{1+\partial \ln V_{c}/\partial \ln R}\,F(\zeta,\,\Phi_{1})
\label{eqn:inflow}
\ee
where $F(\zeta,\,\Phi_{1})=f(\zeta)\,S(\zeta,\Phi_{1},\omega)$ includes a weak, non-linear function 
of the induced motions from the perturbation ($f(\zeta)$ where 
$\zeta\equiv \partial R_{1}/\partial R$; with $1/2<f(\zeta)<2$ for all $\zeta$) 
and the phase function $S$ that depends on the relative phases of 
the overdensity/potential amplitude and radial motions/shock locations 
($-1<S<1$), which determines whether the gas is inflowing ($\dot{M}<0$) or 
outflowing ($\dot{M}>0$). 
We show the resulting maximum $|\dot{M}|$, 
given the same normalization condition. 

We see many of the same trends as in Figure~\ref{fig:eigenvalues}, in greater detail. 
With increasing $\beta$, the growth rates (and pattern speeds, 
$|\Omega_{p}|\sim|\gamma|$) decrease, but the dynamic range covered by the 
modes increases weakly. 
The latter simply comes from the fact that the softening 
suppresses local modes and increases the dynamic range over which the disk 
is in effective contact. 
The decrease in $\gamma$ with $\beta$ is 
especially striking if one considers the fastest-growing modes 
(mostly local, therefore not plotted here). 
The maximum inflow rates 
scale weakly, decreasing at high-$\beta$ as expected due to the increased stiffness.
At high $\beta$ we see the increased prominence of 
retrograde modes as in Figure~\ref{fig:eigenvalues}, and in fact these dominate the 
high-$\mathcal{R}$ modes. Recall, at small radii the 
WKB dispersion relation reduces to Equation~\ref{eqn:slowmode.dispersion.stellar}, 
which maximizes $\omega$ at $|kR|=1/\beta$, giving for the power-law disk
$\omega_{\rm max}=(e^{-1}\beta^{-1} - \alpha(2-\eta))\pi\,G\,\Sigma/\Omega\,R$. 
Thus, when $\beta>1/(e\,\alpha\,(2-\eta))$, $\omega_{\rm max}<0$, 
and only retrograde modes are supported at small radii. As a consequence, 
prograde modes, although they exist, will tend to cover a relatively smaller 
dynamic range. 

We see even at fixed $M_{d}$, the trend that $\gamma$ increases with 
increasing slope $\eta$. The pattern speed $\Omega_{p}$ 
increases with $\gamma$. 
The inflow rates increase towards steeper $\eta$, which simply comes from the 
fact that there is proportionally more disk mass at small $R$. 
There is also significant change in dynamic range. From $\eta\approx0-0.5$, 
$\mathcal{R}$ increases by a factor of $2-3$ with $\eta$, to a peak at 
$\eta\approx1$, and then declines. This comes from the behavior discussed in \S~\ref{sec:structure}, 
wherein modes at $\eta<1/2$ cannot propagate efficiently to $R\rightarrow0$, 
while modes at $\eta>1$ can technically propagate but do so with declining 
efficiency \citep[see][]{hopkins:cusp.slopes}. 
The effect is large: mode dynamic range goes from 
$\sim0.1$\,dex to $\sim2$\,dex in the unstable modes or $0.5$\,dex to $5\,$dex 
in the stable modes. 

Changing the disk mass $M_{d}/M_{\rm BH}$, we also obtain a number of 
interesting trends. For the chosen $\eta$ and $\beta$, unstable modes first appear 
at $M_{d}/M_{\rm BH}\sim0.1$; the criterion depends somewhat on these 
quantities, but in general we find it agrees more or less with our WKB instability 
criterion in Equation~\ref{eqn:instab.criterion}. 
At smaller $M_{d}/M_{\rm BH}$, there are only stable modes -- these persist 
at arbitrarily small $M_{d}/M_{\rm BH}$ and can still drive 
quite significant eccentricities and inflow rates, and at sufficiently 
small $M_{d}/M_{\rm BH}$ their properties can 
be well-approximated by the derivations in \citet{tremaine:slow.keplerian.modes}. 
Once instability arises, there is only a weak dependence of the growth rate 
and $\Omega_{p}$ on 
$M_{d}/M_{\rm BH}$, however (it is largely determined by $\Omega$ at 
the radius where $M_{d}$ reaches this threshold mass). 
Inflow rates increase approximately in proportion to the disk mass.

\subsubsection{Mode Structure}

Figures~\ref{fig:m1.1}-\ref{fig:m1.5} illustrate some of the 
eigenvectors (normal modes) for a range of eigenvalues 
$\omega$, for representative choices of $\eta$ and $\beta$. 
We focus on the global modes; there are of course a wide spectrum of local modes available 
on large scales of the self-gravitating disk. 
In each Figure, we plot the mode amplitude $|a(R)|$, as well as 
${\rm Re}(\Sigma_{a}/\Sigma) = |a(R)|\,\exp{\{i\,\int^{r}k\,dr \}}$, 
which shows the winding and corregations of the modes. 
The surface density perturbations generically experience sharp gradients and reflect 
off the OLR (Re$(\omega)=\Omega+\kappa$, $r\sim$\,a few), 
although in a some cases the $\Sigma_{a}\rightarrow0$ earlier, at COR (a typical 
factor $\sim3$ smaller radius). 

We also plot the vector radial perturbation, defined for simplicity 
here as just $R_{a}/R$ ($|R_{a}/R|$ is equivalent to the standard scalar eccentricity 
in the nearly-Keplerian regime; also in this regime 
$|R_{a}/R| \approx |v_{a}/V_{c}|$ where the velocity perturbation $v_{1,\,R} = dR_{1}/dt$). 
This clearly reflects the $\Sigma$ perturbation. 

We also show the magnitude of the radial perturbation, 
$|R_{1}|/R$ (defined above) 
along with the magnitude of its dimensionless derivative, 
$|d\,R_{1}/d\,R|$, and 
the same for the azimuthal perturbation $\phi_{1}$. 
If these quantities exceed unity, there are orbit crossings, 
therefore shocks and gas dissipation. 
Note that orbit crossings can occur, in principle, 
even if these quantities do not exceed unity - this is only an approximate 
guideline. In fact, \citet{hopkins:inflow.analytics} show that 
in a disk with finite gas sound speed $c_{s}$, 
there can be gas collisions and shocks, even 
where there are no orbit crossings. The requirement for 
dissipative encounters is that gas streams 
moving in the potential of the mode be compressed by the 
torques from the collisionless component at a velocity 
greater than $c_{s}$. This condition can be written 
\be
\xi \equiv \frac{2\pi\,Q}{1-|d R_{1}/d R|}\,\frac{1}{\kappa}\,\left( 
\left| \frac{\partial v_{1,\,R}}{\partial R} \right|^{2} + 
\left| \frac{\partial v_{1,\,\phi}}{\partial R} \right|^{2} \right)^{1/2}
\ge 1
\label{eqn:shocks}
\ee
where $Q=c_{s}\,\kappa/\pi\,G\,\Sigma$. 
This depends on the gas sound speed $c_{s}$, so is not determined 
in the collisionless models here, but if we assume the gas in these disks 
has $c_{s}\sim \sigma_{z}$, then we can determine $\xi$. 
The results are plotted; at small radii especially, $\xi\gg1$ is common -- 
for disks with finite $c_{s}\sim \sigma_{z}$ in their gas, shocks 
will be common (owing to the large $\sim V_{c}$ radial motions of the 
eccentric modes) at small radii, even when formal orbit crossings 
are rare. 

Finally we show the induced inflow/outflow rates, using the 
scaling in Equation~\ref{eqn:inflow} to determine the gas inflow 
response to the stellar+BH potential. Because of our choice of 
units, these inflow rates are in units of 
$f_{\rm gas}\,M_{\rm BH}\,(G\,M_{\rm BH}/R_{0}^{3})^{1/2}$, 
where $f_{\rm gas}=M_{\rm gas}/M_{\ast}$ is the gas mass fraction in 
the disk. The inflow rates can be quite large for the systems of interest -- 
given typical numbers, a rate of $\sim10^{-4}$ in these plots 
corresponds to $\sim0.1-1\,M_{\sun}\,yr^{-1}$ accretion rates onto a 
supermassive BH, or $\sim10^{-5}\,M_{\sun}\,yr^{-1}$ onto a star. 

Now, consider specifically Figures~\ref{fig:m1.1}-\ref{fig:m1.2}, which 
illustrate modes with different frequencies and radial wavenumbers, 
for a fixed disk system. 
We see several of the behaviors discussed in \S~\ref{sec:sims}: 
the modes are global in dynamic range, but can have sizeable $|kR|$ and a 
large number of nodes. 
Since the slopes here are steeper than or 
equal to $\eta=1/2$, the modes can exist down to $r\rightarrow0$. 
Modes with smaller pattern speed extend to larger radii (the OLR moves out) 
and can sustain their amplitude over a larger dynamic range, 
while modes with Re$(\omega)\gg1$ are not supported at moderate to 
large radii. Modes with large negative pattern speed are not seen; but 
those with moderate pattern speed are present are are among the 
most long-wavelength modes. This behavior is similar to that of the $g$-modes in 
\citet{tremaine:slow.keplerian.modes} -- 
however, that work concluded that when the disk mass is negligible 
compared to the BH mass, such modes cannot be supported in an 
isolated BH+disk system; here, we find that moderate disk-to-BH mass ratios 
allow for their existence. 

The induced velocity perturbation and eccentricities are large, and 
reflect the mode density structure. 
Where present, orbit crossings tend to occur between the COR and OLR 
($R\sim R_{0}$), and at the smallest radii. 
However, one is rarely in the strong orbit-crossing regime -- there seems to 
almost be a maximum $|d R_{1}/d R|\sim1$. This is easy to understand from 
the WKB analysis: recall, we showed that in the nearly-Keplerian, 
low disk mass limit, $|v_{r}/V_{c}|\approx |a|/|kR|$. 
Since we are nearly Keplerian, it is trivial to show that correspondingly 
$|R_{1}/R| \approx |v_{r}/V_{c}|$. And in the WKB limit, $d/dR=i\,k$, so 
$|d R_{1}/d R| \approx |a|$. 
Since $|a|<1$ always, we do not expect to see dramatic orbit crossings. 
However, the criteria for shocks, from \citet{hopkins:inflow.analytics}, 
is satisfied over a wide range of radii with $\xi\gg 1$. This occurs because
the perturbation velocities are a fixed fraction $\sim|a|/|kR|$ of $V_{c}$, 
and so diverge at small radii -- thus even without formal orbit crossings, 
gas is being compressed in the mode at velocities $\gg c_{s}$ at 
small radii, generating shocks and dissipation. 

This, correspondingly, allows for the inflow rates calculated (wherever $\xi\gtrsim1$). 
The inflow rates drop towards $R\rightarrow0$, as seen in 
hydrodynamic simulations (see Figures~5 \&\ 12 in \citealt{hopkins:zoom.sims}). 
But given the choice of units here, they are still quite large at 
radii $\sim 10^{-3}-10^{-2}\,R_{0}$, where for the appropriate choice of 
$\Omega_{p}$ and $\eta$, they often asymptote to approximately 
constant $\dot{M}(R)$. Recall again the units used: 
if the value in Figure~\ref{fig:m1.1} is $\tilde{m}$, then the implied 
Eddington ratio of the BH is 
$\dot{M}/\dot{M}_{\rm Edd} \approx 
892\,\tilde{m}\,f_{\rm gas}\,(M_{\rm BH}/10^{8}\,\msun)^{1/2}\,(R_{0}/10\,{\rm pc})^{-3/2}$, 
or $508\,\tilde{m}\,f_{\rm gas}\,(M_{\rm BH}/10^{8}\,\msun)^{-1/4}\,
(\Sigma_{0}/10^{5}\,\msun\,{\rm pc^{-2}})^{3/4}$. 
So the implied accretion rates at these radii are about Eddington, for a 
mode of near-maximal strength. 
This is important, since these are the radii where 
we expect the viscous disk to take over -- inside such a radius, the 
ratio of the disk to BH mass is as low as $\sim10^{-5}$, so $Q$ becomes 
extremely large and we expect star formation 
(and thus the role of the collisionless component) 
to be inefficient.  For a circum-stellar disk, these radii approach 
the stellar radius itself.

In Figure~\ref{fig:m1.3}, we consider how the modes depend on 
the disk mass profile slope $\eta$. When the slope is shallow, the modes are confined 
to larger radii; at $\eta\sim0.5-1.0$ mode amplitudes propagate to $R\rightarrow0$; 
and at much larger $\eta$ the relative amplitude can be damped at smaller $R$ 
making orbit crossings more concentrated in $R$. 
%to a 
%moderate range of radii, with the induced eccentricity, 
%mode crossings, and inflow sharply restricted there as well 
%(inside this inner cutoff, in fact, there is outflow towards the mode). 
%At the value $\eta=1/2$, the modes can suddenly propagate to 
%arbitrarily small $R$. Going to somewhat larger $\eta=0.7-0.8$, 
%the structure of the modes is quite similar. 
%At $\eta$ around unity we see maximally efficient inflows. 
%But at $\eta$ significantly larger than unity, we see the 
%suppression effect discussed in \S~\ref{sec:wkb}. The mode 
%formally has non-zero amplitude to arbitrarily small $R$, but 
%the efficiency of inwards-propagation of the eccentricity 
%is rapidly suppressed. As a consequence, the locations of orbit 
%crossings becomes much more concentrated in radius. 
%Together, these effects suppress further inflow to small $R$ 
%(until the profile becomes again slightly more shallow). 

In Figure~\ref{fig:m1.4}, we compare the mode structure 
(at otherwise similar properties) to the gravitational softening 
or disk thickness. 
At {\em fixed} $\omega$ and other disk properties, 
the modes in colder disks are higher-$k$ 
(more tightly wound). This is also evident in the simulations in 
\citet{hopkins:zoom.sims} (see their Figures~2 \&\ 4). 
This is because, for almost all the modes of interest, the 
mode has at least some contributions from the short-branch 
regime -- they are analogous to the ``p-modes'' in 
\citet{tremaine:slow.keplerian.modes}, both in that 
the pressure/softening effects are non-negligible, in that 
the characteristic modes are trailing (the $kR>0$ branch of the 
p-modes remains $kR>0$ 
after refraction), and in that they have positive (prograde) pattern speeds. 
Because these modes depend on some non-zero $\beta$, 
as the modes themselves heat up the disk when they go non-linear (ultimately 
stabilizing it at some $Q$ threshold), they can become more global. This has the effect 
(at fixed mode amplitude) of 
actually increasing the efficiency of the modes at driving large eccentricity and 
inflows to small radii, although the mode growth rates are lower. 
However, most of this difference depending on disk thickness is concentrated 
at small softening; 
once moderate disk thickness $\gtrsim0.05-0.1$ is reached, 
the effect of the modes is actually fairly weakly dependent on the thickness. 

Figure~\ref{fig:m1.5} compares the mode structure at 
fixed $\eta$ and $\beta$ but varying $M_{\rm disk}$ (and correspondingly $a$). 
At low $M_{\rm disk}/M_{\rm BH}\sim0.1-0.3$,  the first unstable modes to appear 
are, unsurprisingly, the maximally unstable modes with large 
$|kR|\approx1/\beta$. 
As $M_{\rm disk}$ increases, the spectrum of unstable modes expands to 
include longer-wavelength modes, and by $M_{\rm disk}\gtrsim M_{\rm BH}$ includes 
very global modes. 
Once $M_{\rm disk}\gg M_{\rm BH}$, the structure at the radii we are interested in -- i.e.\ in the 
quasi-Keplerian regime {\em inside} $\sim R_{0}$, is essentially independent of 
$M_{\rm disk}$ (it is identical to the case of an infinite power-law disk). Of course, 
there will in this limit be other modes at larger radii that are simply standard disk modes, 
but we are not interested in these behaviors.

\section{Comparison with Simulations}
\label{sec:sims}

Our analysis thus far is restricted to the linear regime. To see whether our key conclusions are 
robust in the non-linear regime, with gas+stellar systems (albeit still 
stellar-dominated), in the presence of inflow, star formation, feedback, and a non-trivial 
potential, we briefly compare to the mode structure in the simulations of self-consistently 
formed nuclear stellar disks from \citet{hopkins:m31.disk}.

\begin{figure}
    \centering
    \scaleup
    \plotone{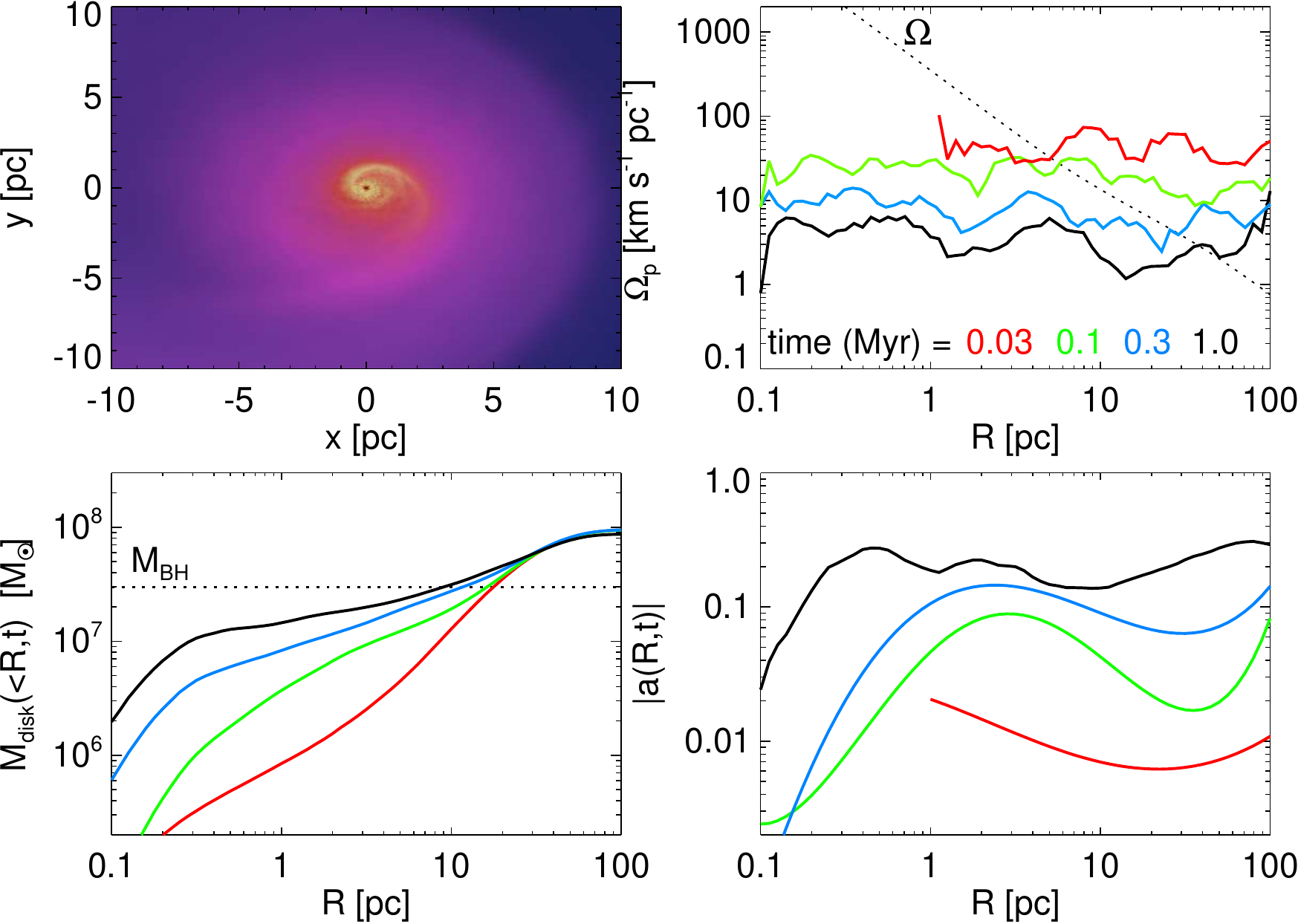}
    %\plotone{f1.pdf}
    \caption{Nuclear mode origins, illustrated in a case study of a single, 
    typical zoom-in simulation of nuclear scales showing the formation of the 
    nuclear $m=1$ mode that ultimately dominates accretion. 
    {\em Top Left:} Image of the (saturated) mode. Brightness shows 
    gas surface density, color (blue to yellow) encodes the specific SFR. 
    {\em Top Right:} Best-fit mode pattern speed $\Omega_{p}$ (solid lines) in the simulation 
    at each of several different times. 
    Dotted line shows the circular velocity $\Omega(R)$. 
    Modes are not plotted where they have no measurable amplitude. 
    {\em Bottom Right:} Enclosed disk mass $M_{d}(<R)$, at the same times. 
    Horizontal dotted line shows the BH mass. 
    {\em Bottom Right:} Mode amplitude $|a|$ in the stellar disk. 
    The mode originates where $M_{d}\sim M_{\rm BH}$, as a locally fast mode 
    ($\Omega_{p}\sim\Omega(R)$). It then propagates inwards, becoming a slow 
    mode at smaller $R$. With time, the inflows generated move the inner instability 
    radius inward, and the slowdown of the mode and change of mass profile 
    shape (from those inflows) allow the mode to propagate to $r\rightarrow0$, 
    as shown. The stellar mode can be sustained for very long times, and 
    at later times drives the gas mode (because stars dominate the mass). 
    \label{fig:nuclear.mode.origins}}
\end{figure}

Figure~\ref{fig:nuclear.mode.origins} illustrates this in a typical 
nuclear scale simulation. We plot the 
enclosed disk $M_{d}(<R)$ and BH mass, 
mode pattern speed $\Omega_{p}$ (and circular speed $\Omega$), 
and mode amplitudes, as a function of time 
in a system that is initially smooth (i.e.\ has no $m=1$ perturbation). 
The mode first appears at some 
radii $\sim R_{\rm crit}$, where $M_{d}/M_{\rm BH}\sim1$, with 
$\Omega_{p}\sim\Omega(R_{\rm crit})$.
At early times, the inner disk profile is quite shallow (or even hollow), because 
no inflow has yet reached the center; we see the resulting 
cutoff in the range of the mode at small radius. 

Two things work to 
push this range inwards. First, 
the mode slows down at early times, seen in Figure~\ref{fig:nuclear.mode.origins}. 
This occurs both via angular momentum exchange with the 
bulge and disk at somewhat larger ($\sim100\,$pc) radii (a 
resonant process not included 
in our analysis), and via direct carrying 
of some of the angular momentum in wavepackets in the gas after reflection off the inner radius 
above (through the OLR, apparent in the WKB treatment). 
This slowdown occurs while the system is in transition from the overstable 
growth phase to the nonlinear, quasi-steady state. 
It halts once the $\Omega_{p}$ is significantly below 
$\Omega(R_{\rm crit})$; in these simulations 
the mode pattern speeds tend to stabilize at 
values $\sim1-5\,{\rm km\,s^{-1}\,pc^{-1}}$. 
The process is analogous to the well-studied process of bar slowdown 
in unstable disks (although obviously with the bulge replacing the halo, which is dynamically 
irrelevant here), and we refer to those studies for further details 
\citep{weinberg:bar.dynfric,
hernquistweinberg92,athanassoula:bar.slowdown,
martinezvalpuesta:recurrent.buckling}, 
although there can be a non-trivial contribution to slowdown from 
the motion of the BH itself 
(damping via scattering off the background stars), an effect not included 
in analytic treatment \citep[compare e.g.][]{adams89:eccentric.instab.in.keplerian.disks,
shu:gas.disk.bar.tscale}. 
As $\Omega_{p}$ decreases, the barrier in Equation~\ref{eqn:slowmode.dispersion.stellar} 
likewise decreases, allowing the mode to strengthen and propagate inwards. 

Second, the mode generates substantial gas inflows, which 
have three important effects. They lead to both gas mass available for further inflow, 
and form stars, which (discussed below) can sustain a long-lived mode at each radius. 
They directly move the $\tilde{f}_{d}\sim0.1$ 
radius inwards, allowing the simple instability region to come closer to the BH. 
They also, correspondingly, steepen the surface density profile. 
Inflows will continuously increase the slope as long as inflow is ``stalled'' at any radius, 
and this is clearly reflected in Figure~\ref{fig:nuclear.mode.origins} as the enclosed disk 
mass at $R<R_{\rm crit}$ rises rapidly with time. 
%The surface density profile 
%rapidly steepens to something similar to that seen in \citet{hopkins:zoom.sims} 
%and \citet{hopkins:m31.disk} (see their Figures~10 and A1, respectively) -- a power-law slope with 
%$\Sigma_{d} \sim R^{-\eta}$ with, eventually $\eta\gtrsim1/2$ ($\eta\sim1/2-1$). This is 
%clear from the behavior seen in $M_{d}(<R)$. 
At this point, because of significant eccentricity at all radii, 
$v_{r}\sim V_{c}$ and $R_{1}\sim R_{0}$, and orbit crossings can occur at 
all radii where the mode is supported. Thus, strong inflows can be sustained 
in systems such as that in Figure~\ref{fig:nuclear.mode.origins} 
down to radii where the BH completely dominates the potential.

\begin{figure}
    \centering
    \scaleup
    \plotone{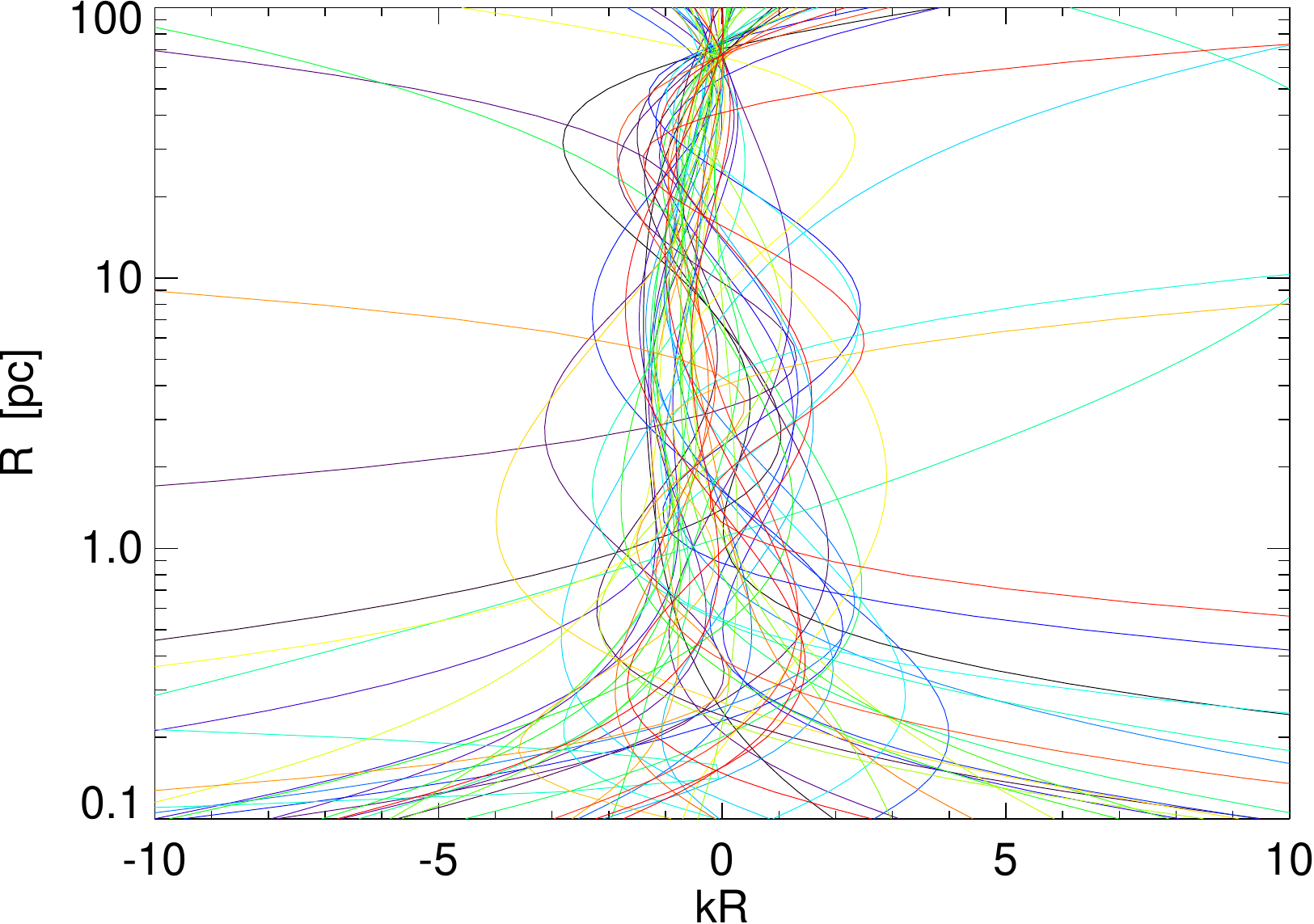}
    %\plotone{f1.pdf}
    \caption{Nuclear mode structure. The radial wavenumber 
    $k$ versus radius $R$, fitted to modes at times near the peak of inflow for 
    each of our nuclear-scale simulations. 
    The modes are clearly global ($|kR|\sim1$) at most radii. 
    Some remain low-$k$ to small $R$, but others wind up 
    (although they can still propagate at finite $k$ to $R\rightarrow0$). 
    Most wind up near $\sim100\,$pc, the effective OLR (but behavior 
    here is complicated by interaction with the outer modes). 
    \label{fig:nuclear.mode.structure}}
\end{figure}

The system quickly reaches a quasi-equilibrium state. The 
$m=1$ mode has been excited down to $R=0$; the mode pattern speeds 
stabilize and the mode amplitudes saturate. 
Figure~\ref{fig:nuclear.mode.structure} shows the structure of the modes, 
specifically the relation between $R$ and $kR$, in different simulations at 
different times. For each simulation snapshot shown, we fit the density distribution over 
the radii plotted to an asymmetric $m=1$ mode of the form 
$\Sigma_{1}=|a(r)|\,\Sigma_{0}(r)\,\exp{\{i\,(\int^{r}\,k(r)\,{d}r + \omega\,t - \phi   ) \}}$, where 
$\Sigma_{0}$ is the aximuthally-averaged surface density and 
we allow $|a(r)|$ and $|k(r)|$ to vary as fifth-order polynomials in 
$\log{(r)}$ (the exact order makes little difference, though much higher order terms are 
noise-dominated). We then plot the instantaneous $R$ and $kR$ for each wave. 
Much of the behavior described above is evident: the modes are global 
$|kR|\sim1$ at most radii -- especially their origin -- but many wind up at small 
$R\lesssim 1\,$pc and 
large $R\gtrsim100\,$pc. 
We see both short and long branches appear in our 
simulations, but the short branch is more common -- this appears to be set as a consequence 
of the ``initial conditions'' set by the earlier mode behavior, 
since in the earlier times while the mode moves to smaller $R$, 
it must wind up (given a shallow ``pre-inflow'' initial profile. 
At the largest radii, the real behavior is complicated by the 
interaction with the modes from the larger-scale inflows. At small radii, the behavior 
is anticipated by our arguments above. 

%That there are some significant differences should not be surprising: not only 
%do our simulations include full non-linear effects, but at radii $\gtrsim1-10\,$pc 
%the structure is more complex than a pure power-law disk. 
%Most important, the softening $\beta$ is not constant with radius in the simulations. 
%In general however, there is reasonable agreement between our analytic expectations and 
%the behavior in simulations. 

\section{Discussion and Conclusions}
\label{sec:discussion}

Lopsided or eccentric disk ($m=1$) modes in quasi-Keplerian potentials 
are of considerable interest 
for a wide range of topics in astrophysics. 
Here, we have discussed how these modes originate in 
collisionless disks, with particular focus on the case of 
mostly stellar disks surrounding a supermassive BH. 
We discuss the origin, structural properties, 
propagation efficiencies, and resonance structure of 
such modes, 
comparing the simple analytic mode 
structure obtained from the WKB approximation, 
exact numerical solutions for the global linear mode structure in idealized disk+BH 
systems, and a full treatment of hydrodynamic 
simulations with gas+stars+BHs, star formation, 
and stellar feedback. 

The solutions here clearly demonstrate that global, over-stable modes with 
linear growth rates $\gamma>0$ 
are, in fact, normal modes of a BH+disk system and can propagate to 
$r\rightarrow0$. 
In other words, our growing modes can be growing even at 
very small radii near the BH, where the enclosed 
disk mass is $\ll M_{\rm BH}$ (here as low as $10^{-5}\,M_{\rm BH}$). 
How can we reconcile this with the demonstration in 
\citet{tremaine:slow.keplerian.modes} that slow modes 
in nearly-Keplerian disks are stable?
The key is that the modes are {\em global}, and the disk 
is not {\em everywhere} quasi-Keplerian (nor is the mode 
everywhere a slow mode). 
If we enforce a cutoff to the disk mass profile or mode 
amplitude such that the regime of the mode is entirely 
inside a radius where 
$M_{d}(<R)\ll M_{\rm BH}$, then we indeed recover 
these previous results and find that growing modes 
are not permitted. But if the disk extends to sufficiently large 
radii such that somewhere, 
$M_{d}(<R)\gtrsim(h/R)\,M_{\rm BH}$ (the canonical 
self-gravity criterion), then at this radius instability is possible. And 
because the modes are global, the mode potential 
$\Phi_{a}$ at small radii $r\rightarrow0$ can still have non-trivial 
contributions from radii where the mode is not slow and the 
potential is not Keplerian. As such, 
it is at least in principle possible to support growing modes at small radii. 

Physically, the correct interpretation is that of the eccentricity 
propagating inwards, as discussed in \S~\ref{sec:structure}. 
In \citet{tremaine:slow.keplerian.modes}, it is noted that the 
non-axisymmetric potential $\Phi_{a}$ does not have to be 
self-generating; it can be imposed as some external $\Phi_{e}$. 
The disk response at small radii is linear in $\Phi_{e}$ -- thus, 
if it has some complex $\omega$, so will the response. 
Here, the effective $\Phi_{e}$ is generated on large scales by 
the self-generating instability, where the disk is self-gravitating 
and the potential is only weakly Keplerian. The inner 
parts of the disk are stable according to the definition of 
\citet{tremaine:slow.keplerian.modes}; their behavior here is their 
linear response to the growing $\Phi_{e}$ imposed from where 
the disk is self-gravitating. 

The $m=1$ mode is special in this respect, in a quasi-Keplerian 
potential, because the system is near-resonance ($\Omega\approx\kappa$), 
so that the excited eccentricity at small radii is independent of the local 
ratio of disk to BH mass. All other (higher-$m$) modes may 
exist, but their propagation efficiencies and ability to induce large 
eccentricities (hence shocks, dissipation, and inflow) 
will be strongly  suppressed by factors $\sim(M_{d}(<R)/M_{\rm BH})$ 
at small radii. 
%Physically, since the natural orbits of stars in a Keplerian 
%potential are ellipses, the outer (growing) part of the mode need only 
%align these orbits (``trapping'' systems in resonance), not strongly distort the 
%initial orbits. 

Such modes appear as 
fast modes ($\Omega_{p}\sim\Omega(R_{\rm crit})$) at the $R_{\rm crit}$ 
where the disk is moderately self-gravitating 
($M_{d}(<R)\gtrsim (h/R)\,M_{\rm BH}$ for local 
modes; for global modes, the criterion depends 
on the exact mass profile but is approximately 
$M_{d}(<R)\gtrsim0.1\,M_{\rm BH}$). 
In this sense they are analogous to any unstable disk mode. 
The stellar waves are bounded by an OLR at a radius typically a factor 
$\sim$a couple in radius beyond co-rotation, although gaseous waves can move through 
this resonance. For this reason, growth rates of modes in pure fluid disks 
with a hard ``edge'' depend on reflection off that edge, and as such are quite sensitive 
to the boundary conditions 
\citep{shu:gas.disk.bar.tscale,
ostriker:eccentric.waves.via.forcing,
nelson:1998.circumstellar.disk.instabilities}; but in gas+stellar systems or 
systems with a more smooth outer disk, the modes can refract and 
self-amplify without much dependence on the outer disk properties. 
If this represents the ``first generation'' of inflows, the inner 
disk profile may be shallow or hollow, so the mode may not be supported 
down to $r\rightarrow0$ and will reflect off of 
a boundary at a factor of several smaller radii. But the mass profile 
will subsequently steepen owing to these inflows, until propagation to 
$R\rightarrow0$ is possible. 
%However, between this inner boundary and co-rotation ($R\sim R_{\rm crit}$) 
%they generate inflows, which will steepen the mass profile near that 
%boundary. 
%As the profile steepens, the waves can propagate further inwards, 
%until a critical slope is reached -- for a power-law disk $\Sigma\propto R^{-\eta}$, 
%this is at $\eta=1/2$ -- at which point the waves can propagate 
%(and continue inflow) to $R=0$.
 
Meanwhile, non-linear effects (seen in simulations) may have slowed down the pattern speed by a 
factor of a few. These include exchanges of angular momentum between 
the inner and outer disk/bulge/halo, and direct carrying of angular momentum 
in the induced gaseous waves (if present). 
%, which during the phase where they reflect off a minimum 
%$R$ will ``bounce back'' carrying angular momentum outwards and 
%can propagate and carry that angular momentum through the OLR. 
The slowdown of the mode moves the OLR outwards and the inner 
radius inwards, and allows for more efficient propagation of the eccentricity 
and mass inflows. Once the critical slope is reached, and most of the 
gas mass has turned into stars, the mode reaches a quasi-equilibrium 
state. Since the stellar mode is fully bounded, and there is no buildup of 
mass at an intermediate radius, 
the pattern speed becomes nearly static, and the mass profile shape evolution 
slows down considerably. Large inflows are maintained to small $R$ as long 
as gas is available, with a rate that scales roughly as $\dot{M}/\dot{M}_{\rm Edd}\sim |a|$, 
the mode amplitude.

In terms of structural properties, the modes of interest are global, with 
$|kR|\lesssim$ a couple over most of their dynamic range (a factor 
of $> 10^{4}$ in radius.
Indeed, in simulations, the modes seen are very global 
$|kR|\ll 1$ (essentially a pure elliptical/lopsided disk) over a significant 
dynamic range, but on sub-pc scales ($\sim0.1-0.5\,$pc), 
they can wind up into tighter spirals. Because they are low-$k$ over 
most of the dynamic range, 
their structure is relatively insensitive to the coldness of the disk, for 
realistic values, and in extremely cold systems other effects will 
tend to heat the disks to moderate values quickly. And growing 
modes can be supported even in disks with large thickness 
$h/R\gtrsim0.3$. In fact, as the thickness (or $Q$) increases, the mode growth 
rates do decrease, but most normal modes become more global, increasing 
their ability to pump up eccentricities and drive large inflow rates. 

With sufficiently large mode amplitudes, there will be orbit 
crossings at many points around the elliptical orbit, but one is generally in the 
marginal orbit-crossing regime (in which there is $\sim1$ orbit-crossing 
per orbit, near the axis of eccentricity). However, allowing for some 
gas in the disk with finite sound speed, the modes can easily drive 
shocks and dissipation at most radii where they have significant amplitude, even 
when there are no formal orbit crossings. 
The resulting inflow rates in this regime are treated in 
\citet{hopkins:inflow.analytics}. Using the scalings therein, we estimate the 
inflow rates generated by these systems. For characteristic 
BH masses and radii of influence, these can be very large -- 
corresponding to accretion at or near the BH 
Eddington limit, for moderate mode amplitudes 
$\gtrsim 0.1$. For plausible mass profile shapes in the ``equilibrium'' range, 
these shocks, dissipation, and inflow rates can be sustained down to 
extremely small radii $\sim10^{-4}-10^{-3}\,R_{\rm BH}$, by which point 
the system has become a traditional viscous accretion disk. 
These modes therefore represent a viable and probably 
very important means of powering large accretion rates onto BHs, from 
$\sim0.01-10\,$pc scales. 

%Although we have usually considered the case of modes around a supermassive 
%BH, our results may be generally applicable in a variety of circumstances. 
%Replacing ``black hole'' with ``star,'' and adopting the appropriate level of sound speed support
%our conclusions regarding gaseous disks can be scaled to some cases of 
%protostellar or protoplanetary disks. 
%Recall, these disks typically have mass ratios relative to the central object 
%within the range discussed here (say $\sim0.01-10\,M_{\ast}$); likewise, approximate 
%power-law profiles of such systems 
%in simulations \citep{laughlin:1994.protostar.disk.instabilities,
%bate:1995.protobinary.accretion.vs.frag,
%nelson:1998.circumstellar.disk.instabilities} 
%and observations 
%\citep{burrows:1996.hh30.disk,zhou:1993.protostellar.collapse,hester:1996.m16.imaging}
%fall within the range explicitly modeled 
%here, as do their pressure profiles. 
%Since the power law disk models here are scale-free, they can be trivially rescaled. 
%However, we caution that an implicit assumption in some of our modeling is that the 
%gas is collisional and governed by an effectively isothermal sound speed (with rapid 
%cooling the typical regime in galactic disks); many of the dynamics of modes 
%will be modified if the cooling time is comparable to the dynamical time 
%\citep[see e.g.][]{gammie:2001.cooling.in.keplerian.disks,
%durisen:2007.grav.instab.in.protoplantary.disks}. 
%Obviously our results should not be applied to these disks 
%when the $m=1$ mode is {\em not} the dominant perturbation.

Although we have usually considered the case of modes around a supermassive 
BH, our conclusions regarding collisionless disks may also 
be applicable in a number of non-galactic regimes. 
For example, a circum-stellar 
disk of planetesimals, or a system with a large density of planets. Comparing 
our results to those in \citet{zakamska:eccentricity.wave.propagation}, who study the efficiency of eccentric 
perturbation propagation in planetary systems, 
we find that most of our conclusions still obtain despite the discrete nature 
of planetary systems, provided we take $\Sigma$ to be the smeared-out (average) surface 
density profile. The mechanics of excitation and characteristic 
frequencies are similar, and 
they show that the efficiency of eccentricity propagation 
decreases for $\eta\gtrsim1$, which we also find. 
If the mass in the collisionless (say, planetary) component of such disks is 
significant compared to the gas mass, and if 
the gaseous component can radiate efficiently when experiencing 
shocks, then our conclusions regarding inflow rates induced by these modes 
should be intact. 
Rescaling our predicted inflow rates to those appropriate for, say, a protostellar 
disk, they can be again quite large -- 
$\sim 10^{-4}\,\msun\,yr^{-1}\,(R[M_{\rm disk}=M_{\ast}]/100\,{\rm au})^{-3/2}$. 
Of course, our speculation regarding the role of star formation 
shaping these profiles on longer time scales should be modified appropriately 
(although there may be some analogies with planet formation in 
such disks). There may also be cases where proto-stellar disks experience 
eccentric perturbations from a collisionless component 
\citep[say, induced 
modes from a binary companion or passages of neighboring 
stars/star-forming cores; see][]{krumholz:2007.rhd.protostar.modes}, 
and these disks typically have mass ratios relative to the central 
object and power-law profiles in the mass range 
explicitly modeled here 
\citep{burrows:1996.hh30.disk,hester:1996.m16.imaging}. Exploring these applications 
in greater detail will be an important subject of future work.

\acknowledgments 
We thank Eliot Quataert and Scott Tremaine for helpful discussions 
in the development of this work. 
Support for PFH was provided by the Miller Institute for Basic Research 
in Science, University of California Berkeley. 
\\

\bibliography{/Users/phopkins/Documents/lars_galaxies/papers/ms}

\end{document}